\documentclass[a4paper,fleqn]{cas-dc}

\usepackage[numbers]{natbib}

\usepackage{amsmath,amssymb,amsfonts}
\usepackage{algorithmic}
\usepackage{graphicx}
\usepackage{textcomp}
\usepackage{verbatim}
\usepackage{bm}
\usepackage{enumerate}
\usepackage{float}  
\usepackage{stfloats}
\usepackage{booktabs}
\usepackage{multirow}
\usepackage{hyperref}
\usepackage[justification=raggedright,singlelinecheck=false]{caption}

\hypersetup{hidelinks,
	colorlinks=true,
	allcolors=black,
	pdfstartview=Fit,
	breaklinks=true}

\def\tsc#1{\csdef{#1}{\textsc{\lowercase{#1}}\xspace}}

\tsc{WGM}
\tsc{QE}
\tsc{EP}
\tsc{PMS}
\tsc{BEC}
\tsc{DE}

\usepackage{tcolorbox}
\tcbset{colback=gray!10!white, colframe=black, boxrule=0.4pt, arc=2pt, 
        left=6pt,right=6pt,top=6pt,bottom=6pt}

\begin{document}
\let\WriteBookmarks\relax
\def\floatpagepagefraction{1}
\def\textpagefraction{.001}

\let\printorcid\relax

\captionsetup[figure]{name={Fig.},labelsep=period,singlelinecheck=off}

\shorttitle{EZhouNet }
\shortauthors{Chu Yun et~al.}

\title [mode = title]{EZhouNet:A framework based on graph neural network and anchor interval for  the respiratory sound event detection.}

\author{Yun Chu}[style=chinese]
\credit{Methodology, Software, Writing - Original draft preparation}

\author{Qiuhao Wang}[style=chinese]
\credit{Investigation}

\author{Enze Zhou}[style=chinese]
\credit{ Software}

\author{Qian Liu}
\cormark[1]
\cortext[cor1]{Corresponding author}
\credit{supervised, fundamental}

\author{Gang Zheng}[style=chinese]
\credit{Hardware and Equipment}

\tcbset{width=\textwidth}
\begin{tcolorbox}

\textbf{Preprint Notice:}  
This is a preprint version of an article submitted to Elsevier.  
The final published version will be available at: \emph{[Journal ]} 
Please note that this preprint may differ from the final published version.
\end{tcolorbox}

\begin{abstract}
Auscultation is a key method for early diagnosis of respiratory and pulmonary diseases, relying on skilled healthcare professionals. However, the process is often subjective, with variability between experts. As a result, numerous deep learning-based automatic classification methods have emerged, most of which focus on respiratory sound classification. In contrast, research on respiratory sound event detection remains limited. Existing sound event detection methods typically rely on frame-level predictions followed by post-processing to generate event-level outputs, making interval boundaries challenging to learn directly. Furthermore, many approaches can only handle fixed-length audio, limiting their applicability to variable-length respiratory sounds. Additionally, the impact of respiratory sound location information on detection performance has not been extensively explored. To address these issues, we propose a graph neural network-based framework with anchor intervals, capable of handling variable-length audio and providing more precise temporal localization for abnormal respiratory sound events. Our method improves both the flexibility and applicability of respiratory sound detection. Experiments on the SPRSound 2024 and HF Lung V1 datasets demonstrate the effectiveness of the proposed approach, and incorporating respiratory position information enhances the discrimination between abnormal  sounds. The reference implementation is available at \url{https://github.com/chumingqian/EzhouNet}.

\end{abstract}


\maketitle

\section{Related work}
In this section, we review related research on graph neural networks and sound event detection tasks in the context of respiratory sounds.

Lang \cite{lang2020graph} et al were pioneers in introducing graph-based techniques to the classification of respiratory sounds, they used spectrograms to represent the relationships between all samples, employing a graph-based semi-supervised learning method to perform binary classification of normal and abnormal lung sounds. Subsequently, in the work \cite{lang2021analysis} they achieved a three-class classification of normal, crackle, and wheeze lung sounds with a limited number of labeled samples and a large set of unlabeled samples. While their introduction of graph-based semi-supervised learning was aimed at handling large-scale unlabeled respiratory sound datasets for classification, they did not extend this work to abnormal event detection in respiratory sounds using graph neural networks.

In \cite{jacome2019convolutional} spectrograms were used as features, and Faster R-CNN was employed as the detection model to identify inspiratory and expiratory phases in respiratory sounds.  \cite{hsu2021benchmarking}  established a public lung sound dataset and evaluated the performance of eight recurrent neural network variants for breath phase and adventitious sound detection. This study confirmed that bidirectional recurrent networks outperformed unidirectional ones and suggested that the inclusion of CNNs improved performance. In subsequent work \cite{hsu2023dual}, they incorporated a tracheal sound dataset to create a mixed dataset of tracheal and lung sounds, training a convolutional recurrent neural network. Experimental results showed that the model trained on multiple domains performed better than when trained on a single domain dataset. In recent work \cite{yeh2024novel}, they completed the task of respiratory sound event detection using the SPR dataset. They first used a convolutional neural network to classify the spectrograms of respiratory sounds, then applied a sound event detection module to identify time regions containing abnormal respiratory sounds, which were later extracted and classified using a trained network.

Inspired by the aforementioned works, we introduce graph neural network (GNN) techniques and the concept of object detection from computer vision into the task of abnormal respiratory sound event detection.

\section{Methodology}

In this section, we will introduce the proposed method, including multi-channel spectrogram generation, the  construction of respiratory sound graph from the spectrogram, the update of node features in the graph, the interval fine-tuning module, and the entire network model architecture.

\subsection{Multichannel Spectrogram Generation}

For the multi-channel spectrogram generation, we first use the  \textbf {Mel spectrogram} and  \textbf {Gamma spectrogram}, both are derived using the Fast Fourier Transform (FFT). For a given fixed-length audio signal $x[n]$ the time-domain signal is transformed into the frequency domain using FFT as  $X[k]_{mel}=\sum_{n=0}^{N-1}x[n]\cdot e^{-{\frac{2\pi i}{N}}k n}$,  where $N$ is the number of FFT points,  $k$  is the frequency bin, and $i$ is the imaginary unit.

Once the frequency-domain signal $X[k]$ is obtained, the  Mel spectrogram  applies a set of Mel filter banks to extract frequency bands. The power spectrum is then log-scaled, resulting in the Mel spectrogram, denoted as $S_0$. The conversion between Mel frequency and Hz frequency is given by $f_{\mathrm{Mel}}=2595\log_{10}\left(1+{\frac{f}{700}}\right)$.

The  Gamma spectrogram  is derived using a gamma-tone filter bank, which enhances different frequency bands of the signal in the frequency domain. The filtered output is log-scaled to generate the Gamma spectrogram, denoted as $S_1$. In contrast to the Mel and Gamma spectrograms, the  Constant-Q Transform (CQT)  converts the signal to the time-frequency domain with a constant Q factor, defined as $Q={\frac{f_{\mathrm{k}}}{\Delta f_{\mathrm{k}}}}$ , where $f_{\mathrm{k} }$  is the center frequency of the $k$-th bin, and   $\Delta f_{\mathrm{k}}$ is its bandwidth. The CQT is computed as $X[k]_{cqt}=\sum_{n=0}^{N-1}x[n]\cdot e^{-{\frac{2\pi i}{Q}}{\frac{k}{f_{s}}}n}$ , the output is then log-scaled to generate the CQT spectrogram, denoted as $S_2$.

These multichannel spectrograms provide diverse time-frequency representations for further analysis. After obtaining the three different types of spectrograms, the multichannel spectrogram features are generated using the following equation  \eqref{1}:

\begin{equation}
S = S_0 \oplus   S_1  \oplus   S_2 \label{1}\\
\end{equation}

here  $S  \in  R^{c\times  m \times n}$ represents the three types of spectrograms. Each spectrogram is an $m \times n$  matrix,  $m$ denotes the frequency dimension, $n$ represents the number of frames, and $c$  corresponds to the number of channels. By using the $\oplus$ operation, the spectrograms are concatenated along the channel dimension, resulting in a three-channel spectrogram feature.

\subsection{Spectrogram to graph construction}

In real-world scenarios, the length of the collected respiratory sounds is often variable. The graph construction module is designed to naturally handle audio with variable lengths by converting the audio spectrogram into a graph-based data structure. This enables the use of graph neural networks to process graphs with varying numbers of nodes, thus allowing the handling of audio with variable lengths.

This part will provide a detailed explanation of how the spectrogram of the audio is converted into a graph data structure, focusing on five aspects: node generation, node label generation, edge construction, graph generation, and batch graph construction.

\subsubsection{ Node Generation}

The multi-channel spectrogram feature $S \in R^{c\times m \times n}$ is first grouped, with each group containing spectrogram information from 5 frames. For convenience, each set of 5 frames is denoted as $x$. This $x$ is then input into the Dynamic Node Generator (DNG), which operates as follows:

First, $x$ undergoes convolution through a Dynamic Conv2D layer. The Dynamic Conv2D consists of $n$ basis convolution kernels, as shown in Equation \eqref{2}.

\begin{equation}
\begin{aligned}
&\alpha = \mathrm{attention2d}(x) \in \mathbb{R}^{B \times n_k \times 1 \times T \times 1} \\
&W_{\mathrm{basis}} \in \mathbb{R}^{n_k \times C_{\mathrm{out}} \times C_{\mathrm{in}} \times k \times k} \\
&W_{\mathrm{agg}} = \mathrm{reshape}\bigl(W_{\mathrm{basis}}\bigr) \in \mathbb{R}^{(n_k C_{\mathrm{out}}) \times C_{\mathrm{in}} \times k \times k} \\
&y = \mathrm{Conv2d}\bigl(x; W_{\mathrm{agg}}\bigr) \in \mathbb{R}^{B \times n_k \times C_{\mathrm{out}} \times T \times F}\\
&\mathrm{DynamicConv}(x) = \sum_{j=1}^{n_k} \alpha_j \, y_j \in \mathbb{R}^{B \times C_{\mathrm{out}} \times T \times F}
\label{2}
\end{aligned}
\end{equation}

Next, the Dynamic Conv2D output is combined with BatchNorm(BN) and ReLU to form a dynamic convolution layer, as depicted in Equation \eqref{3} , with the learnable parameters denoted as $\theta_i$, ${\theta^{(i)}} = (W^{(i)},b^{(i)}) $.

\begin{equation}
\begin{aligned}
&{D_{\theta^{(i)}}(x)}= \mathrm{ReLU}\bigl( \mathrm{BN}\bigl( \mathrm{DynamicConv}(x;W^{(i)},b^{(i)})\bigr) \bigr) \\
& \text{where} \; W^{(i)} \in \mathbb{R}^{C_{\mathrm{out}}^{(i)}\times C_{\mathrm{in}}^{(i)}\times k_i\times k_i}, \quad b^{(i)} \in \mathbb{R}^{C_{\mathrm{out}}^{(i)}} \\
\label{3}
\end{aligned}
\end{equation}

A pooling layer is then added after the dynamic convolution layer to scale the feature map. By combining the dynamic convolution and pooling layers, we obtain the output of the dynamic convolution. This combination pattern is repeated three times, where each layer's input is the output of the previous layer, as shown in Equation  \eqref{4}.

Then the DynamicFeatureExtractor applies three such blocks (with dynamic or standard conv-kernels) each followed by average‐pooling:

\begin{equation}
\begin{aligned}
&x^{(0)} = x_{in} \\
&y^{(i)} = D_{\theta^{(i)}}\bigl(x^{(i-1)}\bigr) \\
&x^{(i)} = \mathrm{AvgPool}_{p_i}\bigl(y^{(i)}\bigr)  \\
& where \quad i=(1,2,3), x \in \mathbb{R}^{B \times n_{\mathrm{in}} \times T \times F}\\
& and \quad p_i =(1,4) \;  \text{for each layer} \\
\label{4}
\end{aligned}
\end{equation}

After obtaining the output from the final dynamic convolution layer, we flatten it along the batch dimension, as shown in Equation \eqref{5}.

\begin{equation}
\begin{aligned}
&f = \mathrm{Flatten}\bigl(x^{(3)}\bigr) \in \mathbb{R}^{B \times \bigl(C_{\mathrm{out}}^{(3)}\,T'F'\bigr)} \\
\label{5}
\end{aligned}
\end{equation}

Finally, as demonstrated in Equation \eqref{6}, the spectrogram features from each group of five frames are mapped to a node feature, denoted as $Node_{i}$.

\begin{equation}
\begin{aligned}
&\mathrm{Node\_i} = \mathrm{ReLU}\Bigl(\mathrm{LayerNorm}\bigl(W_{\mathrm{spec}}\,f + b_{\mathrm{spec}}\bigr)\Bigr) \\
&where \; W_{\mathrm{spec}} \in \mathbb{R}^{d_{\mathrm{node}} \times \bigl(C_{\mathrm{out}}^{(3)}T'F'\bigr)}, \quad b_{\mathrm{spec}} \in \mathbb{R}^{d_{\mathrm{node}}}
\label{6}
\end{aligned}
\end{equation}

\subsubsection{ Node Label  Generation}

We define two types of node labels: the node confidence label, denoted as $Conf_{node}$, and the node category label, denoted as $Y_{node}$.

For the node confidence label, we adopt a soft label approach instead of the traditional hard-coding method. Experimental results show that soft labels improve the model's stability and convergence during training.

Let the chunk’s frame-wise one-hot labels in \eqref{7}

\begin{equation}
  \begin{aligned}
    f_j \;=\;\arg\max_{k\in\{0,\dots,K\}}\;\mathrm{frames}_{j,k}  \qquad j=1,\dots,w
\label{7}
  \end{aligned}
\end{equation}

Define the abnormal mask as follows , Then, we use the ratio of anomalous frames in the current chunk size to the total number of frames in the chunk as the confidence label for the node as shown in \eqref{8} . If all frames in the current chunk are normal, the node confidence label is 0.

\begin{equation}
 \begin{aligned}
  \qquad a_j \;
  &=\;\mathbf{1}\{f_j \neq 0\} \\
  \quad{Conf}_{\mathrm{node}}\;
  &=\;\frac{1}{w}\sum_{j=1}^w a_j
  \quad(\,=0\;\text{if all }a_j=0\,)\\
  \label{8}
\end{aligned}
  \end{equation}

For the node category label, we assign the label corresponding to the most frequent anomalous frame type in the current chunk, as shown in the formula \eqref{9}, where $\ell \in {0,\dots,K-1}$ indexes the four abnormal classes.

\begin{equation}
Y_{\mathrm{node}}
=\begin{cases}
\,-1, & \displaystyle\sum_{j=1}^w a_j = 0\\[8pt]
\displaystyle
\underset{\ell\in\{0,\dots,K-1\}}{\arg\max}\;\bigl|\{j:f_j=\ell+1\}\bigr|,
&\text{otherwise}
\end{cases} 
\label{9}
\end{equation}

If there is a tie between two anomalous types, the label corresponds to the first appearing anomalous frame. This scenario occurs rarely, especially for small chunk sizes. Furthermore, if all frames are normal, the node label is set to -1, meaning that only anomalous categories will be considered in subsequent classification, while normal category are ignored.


\subsubsection{ Edge Construction}

In each sample of $n$ nodes, node $j$ is connected to node $j+1$ in a directed "chain" graph. The edge labels are assigned based on the labels of the connected nodes.  In equation \eqref{10} repre that if one of the nodes is anomalous, the edge label is set to 1, indicating an anomaly.

\begin{equation}
\begin{aligned}
& E^y_{j,j+1} =\;
  \mathbf{1}\bigl\{Y_{\mathrm{node},j}\neq -1 \;\vee\;
Y_{\mathrm{node},j+1}\neq -1\bigr\} \\
& where \; \quad j=1,\dots,n-1
\label{10}
\end{aligned}
\end{equation}

The edge features are generated by projecting the node features of the two connected nodes onto a node-edge space, as shown in the \eqref{11} .

\begin{equation}
  h^{\mathrm{edge}}_{j,j+1}
  \;=\;
  \mathrm{ReLU}\Bigl(
  W_{\mathrm{edge}}
  \bigl[h^{\mathrm{node}}_j \,\Vert\, h^{\mathrm{node}}_{j+1}\bigr]
  + b
  \Bigr)
  \label{11}
\end{equation}

\subsubsection{Graph Generation (per sample)}

After we generate node features, node labels, edge features and edge labels, now we can construct the graph $G_i$ for each sample.

\begin{equation}
\begin{aligned}
  G_i & = \left( V_i,\, E_i,\,
  \underset{\text{Node features}}{X_i},\,
  \underset{\text{Node labels}}{Y_i},\,
  \underset{\text{Edge features}}{E_i^f},\,
  \underset{\text{Edge labels}}{E_i^y} \right) \\
V_i &= \{v_{i,1},\dots,v_{i,n_i}\}\\
E_i &= \{(v_{i,j},v_{i,j+1})\mid j=1,\dots,n_i-1\}\\
\label{12}
\end{aligned}
\end{equation}

As shown in  equation \eqref{12},  where $V_i$  represent the collection of all nodes in the $i-th$ sample.
Each node corresponds to a chunk (e.g., a group of 5 frames) of the input spectrogram, turned into a node feature,and $E_i$ represent the collection of directed edges: each node $j$ connects to node $j+1$, making a chain structure, so every node connects forward to its immediate neighbor, and the edges are directed.

The remaining four variable properties of the graph as shown in formula \eqref{13}.

\begin{equation}
  \begin{aligned}
  \ X_i &=\;
  \begin{bmatrix}
  h^{\mathrm{node}}_{i,1} \\[2pt]
  \vdots \\[2pt]
  h^{\mathrm{node}}_{i,n_i}
  \end{bmatrix}
  \in\mathbb{R}^{n_i\times d}\\
  \quad Y_i &= \bigl[Y_{\mathrm{node},i,1},\dots,Y_{\mathrm{node},i,n_i}\bigr]^\top\\[4pt]
  E^f_i &=\;
  \begin{bmatrix}
  h^{\mathrm{edge}}_{i,1,2} \\[2pt]
  \vdots \\[2pt]
  h^{\mathrm{edge}}_{i,n_i-1,n_i}
  \end{bmatrix}\\
  \quad E^y_i & = \bigl[E^y_{i,1,2},\dots,E^y_{i,n_i-1,n_i}\bigr]^\top
  \label{13}
  \end{aligned}
  \end{equation}

where the 

  - $X_i$: represent the node feature matrix $X_i$ is consist of $h^{Node}_i$,which connects node features and optionally with the  one-hot encoded gender/position.

  - $Y_{i}$.Node labels: 
  
  - $E^f_i,E^y_i$  Edge  features and labels.

\subsubsection{ Batch Graph Construction}

Finally, all sample graphs ${G_i}$ are combined into a batch graph $\mathcal{G}$ by connecting node features and stacking edges with index offsets. This allows a single GNN to traverse the entire batch. 

Let cumulative node counts $ N_0=0,\quad N_i = \sum_{k=1}^i n_k$. The batched feature matrix in \eqref{14} and the batched edge index in \eqref{15}.

\begin{equation}
  X = 
  \begin{bmatrix}
  X_1 \\
  X_2 \\
  \vdots \\
  X_B
  \end{bmatrix}
  \in \mathbb{R}^{(\sum n_i) \times d}
  \label{14}
  \end{equation}

Notably, the $Batch$ object includes the attributes  $edge$  $index$ and $batch$, which indicate the sample each node belongs to and the sample associated with each edge. These attributes ensure that when using a graph neural network for batch processing, node feature updates and message passing occur within each sample, without cross sample message passing or updates.

\begin{equation}
\begin{aligned}
\text{edge\_index} 
&= \left\{ (j + N_{i-1},\, j+1+N_{i-1}) \right\}_{j=1}^{n_i-1}{}_{i=1}^B \in \mathbb{Z}^{2 \times \sum (n_i-1)}\\
\text{batch}  
&=\;
\bigl[\underbrace{1,\dots,1}_{n_1},\underbrace{2,\dots,2}_{n_2},\dots,
\underbrace{B,\dots,B}_{n_B}\bigr]^\top
\label{15}
\end{aligned}   
\end{equation}

In \eqref{16}, batched edge features and labels are simply concatenations of each  $E^f_i$,$E^y_i$.

\begin{equation}
\begin{aligned}
\mathrm{edge\_attr}
&=\;\mathrm{concat}\bigl(E^f_1,\dots,E^f_B\bigr)\\
\mathrm{edge\_y}
&=\;\mathrm{concat}\bigl(E^y_1,\dots,E^y_B\bigr)\\[6pt]
\label{16}
\end{aligned}    
\end{equation}

The Batch object encapsulates the following information in \eqref{17}, until here, we have completed the construction of graph for  the respiratory sound. Next, we will perform node updates and event interval detection based on this graph.

\begin{equation}
  \mathcal{G} : (X,\, \text{edge\_index},\, \text{edge\_attr},\, Y,\, \text{edge\_y})
  \label{17}
  \end{equation}

\subsection{Node Update Module}
The role of the node update module is to update the respiratory  features of each node in the graph. We extend the message-passing framework of graph neural networks by using Graph Attention Networks (GAT) to learn the attention coefficients on the edges.

This allows each respiratory node to aggregate information from its neighboring nodes using the learned, edge-specific weights, enhancing the discriminative power of the updated respiratory sound node features. The core steps are as follows.

\subsubsection{Node Update}

Firstly, each node’s feature vector $\mathbf{h}^{node}_i \in \mathbb{R}^{F_\text{in}}$ is linearly transformed via the Linear Projection in  \eqref{18} :

  \begin{equation}
  $$\mathbf{h}_i = \mathbf{W}\,\mathbf{h}^{node}_i,\quad \mathbf{W}\in\mathbb{R}^{F_\text{out}\times F_\text{in}}$$
  \label{18}
  \end{equation}

 Then, for each edge attribute $\mathbf{e}_{ij}$, an unnormalized attention score $e_{ij}$ is computed used Attention Mechanism as shown in \eqref{19}, After that, we normalize the attention scores ${e}_{ij}$ to achieve the  normalized attention coefficients $\alpha_{ij}$.
  
  \begin{equation}
  \begin{aligned}
  e_{ij} &= \text{LeakyReLU}\bigl(\mathbf{a}^\top[\mathbf{h}_i \parallel \mathbf{h}_j \parallel \mathbf{e}_{ij}]\bigr) \\
  \alpha_{ij} &= \frac{\exp(e_{ij})}{\sum_{k\in \mathcal{N}(i)} \exp(e_{ik})}
  \label{19}
  \end{aligned}
  \end{equation}
  
  where $\mathbf{e}_{ij}$ is the edge attribute with the  dimension of 12, $\parallel$ denotes concatenation, and $\mathbf{a}$ is a learnable vector.

  Then do the neighborhood aggregation,  each node is multiplied by the original node through the normalized attention coefficients, and the result is passed through a nonlinear activation layer, as shown in \eqref{20}.

  \begin{equation}
  $$\mathbf{h}_i' = \sigma\Bigl(\sum_{j\in\mathcal{N}(i)} \alpha_{ij}\,\mathbf{h}_j\Bigr)$$
  \label{20}
  \end{equation}

where $\sigma$ is a nonlinearity (ReLU in the second GAT layer).

We repeat these steps with two  GAT layer to  obtain the updated respiratory sound node features. And then using the temporal positional encoding to enhance node embeddings with temporal context. 
Given an audio input, we already obtained the update node embeddings as we mentioned before, for the convenience of expression, we renote the updated node feature  $ \textbf{h}^{'}_i$ as $\textbf{e}_n$ in \eqref{29}.

\begin{equation}
  \mathbf{E} = [\mathbf{e}_n]_{n=1}^N, \quad \mathbf{e}_n \in \mathbb{R}^D  \\
  \label{29}
\end{equation}

here \(N\) is the number of nodes and \(D\) is the feature dimension, and he normalized node timestamps and the raw node classification logits as shown in \eqref{30}.

\begin{equation}
  \begin{aligned}
  \mathbf{t} = [t_n]_{n=1}^N, \quad t_n \in [0,1] \\
  \mathbf{P} = [\mathbf{p}_n]_{n=1}^N, \quad \mathbf{p}_n \in \mathbb{R}^C
  \label{30}
\end{aligned}
\end{equation}

where \(t_n\) represents the normalized time position of node \(n\), \(C\) is the number of node classes. In addition, normalized anchor intervals at multiple scales \(k\in\{1,2,3\}\) are predefined in \eqref{31}, ${n_k}$ is the number of anchor intervals for scale \(k\).
\begin{equation}
  \{(\alpha_{k,i}, \beta_{k,i})\}_{i=1}^{n_k},
  \quad \alpha_{k,i} < \beta_{k,i}, \quad \alpha,\beta \in [0,1]
  \label{31}
\end{equation}

Since node embeddings \(\mathbf{e}_n\) are extracted independently of their temporal position, we explicitly inject time encoding into the features. In \eqref{32}, we first define a vector of frequencies, then each node’s timestamp is encoded via a sinusoidal function.

\begin{equation}
\begin{aligned}
\boldsymbol{\omega} &= [\omega_d]_{d=1}^D, 
\omega_d = \frac{10\,(d-1)}{D-1} \\
\mathbf{c}_n &= \gamma\,\sin(t_n\,\boldsymbol{\omega}),
\mathbf{x}_n = \mathbf{e}_n + \mathbf{c}_n
\label{32}
\end{aligned}
\end{equation}

where \(\gamma\) is a small scaling factor (e.g., 0.05) to prevent overwhelming the original features. Thus, here \(\mathbf{x}_n\) is the time-aware node embedding.

Later, we use  the updated node features to complete the generation of anchor interval's offsets and node classification.

\subsection{Anchor Interval Refine Module}

The purpose of the anchor interval design is to alleviate the difficulty of predicting event intervals directly by the network. Instead of having the network predict the event interval from scratch, we predefine anchor intervals at three different scales, each with a different duration. This helps the network capture different types of anomalous respiratory sound events. The network then learns the offset based on these anchor intervals. By combining the anchor intervals and their corresponding offsets, we obtain the final predicted intervals.

The working mechanism of the interval refine module will be explained from three aspects: the generation of anchor intervals, the labeling of anchor intervals, and the generation of anchor interval offsets.

\subsubsection{ Anchor-Interval Generation at Three Scales}

Given an input audio of duration $L$, we wish to generate a fixed set of temporal anchor intervals at three different scales (durations).
Let the following variables:
  
$L\in\mathbb{R}^+$: actual audio length (in seconds).
    
$K=3$: number of scales.

In \eqref{21} show the three desired scale durations in seconds.

\begin{equation}
    \mathbf{d} = \left[ d_1,\, d_2,\, d_3 \right] = \left[ 0.5,\, 0.8,\, 1.5 \right]
    \label{21}
\end{equation}

In \eqref{22}, the reference length $L_{\rm avg}=L$ is the audio's actual duration, since each clip is range from 9s to 15s, it's actual length may vary, we need to use normalized scale factors,

\begin{equation}
  \begin{aligned}  
  s_k = \frac{d_k}{L_{\rm avg}}, 
  n_k = \bigl\lfloor N_0 \,w_k\bigr\rfloor,\quad  
        k=1,2,3 \\
  \label{22}  
  \end{aligned}
  \end{equation}

where the weight vector ${w}_k=[0.75,\,2.0,\,0.75]$ is use to control anchor density per scale, the base number of centers $N_0=20$, $n_k$ is the number of centers at scale $k$.
      
Then we  compute the anchor intervals, for each scale $k$:
  
1. First,in \eqref{23} we normalize the  center positions and  half width of the anchor interval.
      
\begin{equation}
  \begin{aligned}  
    c_{k,i} = \frac{i+0.5}{n_k}, \; \tfrac12 s_k,  \quad i=0,1,\dots,n_k-1 \\
  \label{23}  
  \end{aligned}
  \end{equation}

2. Then, in \eqref{24} we  compute each normalized anchor interval's  start/end points, also need clamp  to $[0,1]$:

\begin{equation}
  \begin{aligned}  
    \alpha_{k,i} &= \max\bigl(c_{k,i}-\tfrac12 s_k,\;0\bigr),\\  
    \beta_{k,i} &= \min\bigl(c_{k,i}+\tfrac12 s_k,\;1\bigr).  
  \label{24}  
  \end{aligned}
  \end{equation}

3.After we get the  normalized anchor $(\alpha_{k,i},\beta_{k,i})$, in \eqref{25} we concatenate all $K$ scales and multiply by $L$ to obtain actual-length anchors.

\begin{equation}
  \begin{aligned}  
   A = \{\,(\alpha_{k,i}L,\;\beta_{k,i}L)\;\}_{k=1..K,\;i=0..n_k-1}.  
  \label{25}  
  \end{aligned}
  \end{equation}

\subsubsection{Anchor-Interval labeling Assignment }

After obtaining the anchor intervals for each sample, we also need to assign the corresponding labels to each anchor interval. Each anchor interval label consists of three components: interval confidence, interval category, and interval time localization, denoted as $ (\text{conf}_i,\;\text{cls}_i,\;\text{interval}_i)$.

For interval confidence, we use a soft label approach, consistent with the node confidence type. However, interval confidence is generated based on the Intersection over Union (IoU) between the anchor interval and the ground truth interval. 

If the IoU between an anchor interval and any ground truth interval in the sample is 0, the confidence value of the anchor interval is set to 0, and its corresponding interval category is labeled as -1, indicating a normal interval. The time localization of the interval is set to (0, 0), meaning the network will not optimize for detection of normal intervals during training.

Conversely, if the IoU between an anchor interval and any ground truth interval in the sample is greater than 0, we proceed as follows to obtain the corresponding category label and time localization label for the anchor interval.

For any given respiratory sound sample, let the set of anchor intervals be denoted as $A =\{a_i\}_{i=1}^N$, where each $a_i=[t_i^{\rm s},\,t_i^{\rm e}]$ represents the start and end times of each anchor interval.

Let the set of ground truth intervals for the sample be denoted as $M= \{g_j\}_{j=1}^M$  , where each $g_j=[\tau_j^{\rm s},\,\tau_j^{\rm e}]$ represents the start and end times of each ground truth interval, and $\ell_j\in\mathcal{L}$ denotes the anomaly type for each ground truth interval.

For each prior interval \( a_i \), we calculate the IoU between \( a_i \) and all ground truth  $g_j$ intervals in \eqref{26}:

\begin{equation}
\begin{aligned}  
&\text{Intersection:}\quad  
T_{i,j}^{\rm I}  
= \max\!{\bigl(0,\;  
\min(t_i^{\rm e},\,\tau_j^{\rm e})  
- \max(t_i^{\rm s},\,\tau_j^{\rm s})  
\bigr)}\\  
&\text{Union:}\quad  
T_{i,j}^{\rm U}  
= (t_i^{\rm e}-t_i^{\rm s})  
+(\tau_j^{\rm e}-\tau_j^{\rm s})  
-T_{i,j}^{\rm I},\\  
& \text{IoU}_{i,j} = \frac{T_{i,j}^{\rm I}}{T_{i,j}^{\rm U}+\epsilon}  \quad  \bigl(\epsilon\!\approx\!10^{-6}\bigr) 
\label{26}  
\end{aligned}
\end{equation}


In \eqref{27}, we identify the maximum IoU value for \( a_i \), and define an IoU threshold \( \theta_{iou} \), only if the maximum IoU exceeds this threshold the maximum IoU will be assigned as the confidence for the interval and denoted as \( \gamma_i \), otherwise, the interval will be labeled as normal.

\begin{equation}
  $$ \quad  \gamma_i = \max_{j}\text{IoU}_{i,j},i^* = \arg\max_{j}\text{IoU}_{i,j} 
  \label{27}$$  
\end{equation}

Upon obtaining the maximum IoU value for \( a_i \), we define another variable \( i^* \), which represents the index of the ground truth interval corresponding to the maximum IoU. Using \( i^* \), we retrieve the category label \( C(\ell_{i^*}) \) and time localization \( (\tau_{i^*}^{\rm s}, \tau_{i^*}^{\rm e}) \) for the corresponding ground truth interval, which are then assigned as the category label and time localization label for the anchor interval \( a_i \).

Through this process, we can achieve the confidence, category, and localization labels for each prior interval, which are represented by the  formula \eqref{28}.

\begin{equation} 
\begin{aligned}
(& \text{conf}_i,\;\text{cls}_i,\;\text{interval}_i) = \\ 
&  \begin{cases}  
\bigl(\gamma_i,\;\;C(\ell_{i^*}),\;\;( \tau_{i^*}^{\rm s},\,\tau_{i^*}^{\rm e})\bigr),  
&\text{if }\gamma_i\ge\theta\text{ and }\ell_{i^*}\neq\text{“Normal”}\\  
(0,\;-1,\;(0,0)),  
&\text{otherwise}  
\end{cases}  
\label{28}
\end{aligned}
\end{equation}

where the,
  
$\text{conf}_i \in[0,1]$: soft confidence = best IoU.

$\text{cls}_i\in\{0,1,\dots,C-1\}\cup\{-1\}$: integer class index ($-1$ for background whihc represent the  normal interval ).

$\text{interval}_i$: the matched GT interval or dummy $(0,0)$.

\subsubsection{ Generating Anchor-Interval Offsets}

After we generate a set of candidate anchor intervals at multiple scales, we need to design the Anchor Interval Refine module to adjust their boundaries based on localized node level context. This module design allows the network to learn how local patterns within an anchor should influence its final start/end times. The main steps are the following.

1.Anomaly score extraction via learnable smoothing. To better capture local node anomalies, in  \eqref{33} we  smooth the raw node class logits \(\mathbf{P}\) using a Gaussian-initialized 1D convolution grouped by class channels. This operation allows for learnable smoothing across neighboring nodes. Then, the smoothed logits are passed through a softmax to obtain smoothed class probabilities and anomaly score.

\begin{equation}
\begin{aligned}
\widetilde{\mathbf{P}} &= \mathrm{softmax}\bigl(\mathrm{Conv}_{\rm gauss}(\mathbf{P})\bigr) \\
\quad  s_n &= \widetilde{P}_{n,\,\text{“abnormal”}}
\label{33}
\end{aligned}
\end{equation}
where \(s_n\) represents the anomaly score of node \(n\),  the probability that it belongs to the abnormal class.

2.Selection of nodes per Anchor Interval. For each anchor interval \((\alpha_{k,i}, \beta_{k,i})\) at scale \(k\), we select the nodes whose timestamps fall inside the interval.This is achieved by constructing a binary mask in  \eqref{34}:
 
\begin{equation}
  \begin{aligned}
  M_{k,i,n} &= \mathbf{1}\left[\,
  \alpha_{k,i}L \;\leq\; t_n L \;\leq\; \beta_{k,i}L
  \right] \\
  \quad \mathcal{N}_{k,i} &= \{n \mid M_{k,i,n}=1\}
  \label{34}
\end{aligned}
\end{equation}
here \(L\) is the  length of the timeline. Nodes inside \(\mathcal{N}_{k,i}\) are then gathered for further feature extraction.

3.Local Feature Extraction with GRUs. For each anchor, we treat the selected sequence of node features and anomaly scores as local observations and process them through Gated Recurrent Units (GRUs) to capture sequential dynamics, as shown in \eqref{35}, we use two separate GRUs,  feature GRU encoding and anomaly score GRU encoding. 

- One for the time-aware node embeddings \(\{\mathbf{x}_n\}\).

- One for the anomaly scores \(\{s_n\}\).

\begin{equation}
\begin{aligned}
\mathbf{h}_{k,i} = \mathrm{GRU}_{k}^{\rm feat}\bigl(\{\mathbf{x}_n\}_{n\in\mathcal{N}_{k,i}}\bigr)
\in \mathbb{R}^D \\
a_{k,i} = \mathrm{GRU}_{k}^{\rm anom}\bigl(\{s_n\}_{n\in\mathcal{N}_{k,i}}\bigr)
\in \mathbb{R}
\label{35}
\end{aligned}
\end{equation}

where \(\mathbf{h}_{k,i}\) and \(a_{k,i}\) are the final hidden states summarizing the local context.

4. Construction of Shared Anchor Features. In \eqref{36}, the anchor interval's centers is further described by its geometric information, $center$ and $width$ of the anchor.

\begin{equation}
c_{k,i} = \frac{\alpha_{k,i} + \beta_{k,i}}{2},
\quad
w_{k,i} = \beta_{k,i} - \alpha_{k,i}
\label{36}
\end{equation}

Then, in \eqref{37} we concatenate the GRU outputs and anchor geometry into a shared feature vector denoted as \(\mathbf{z}_{k,i}\).

\begin{equation}
\mathbf{z}_{k,i}
= \Bigl[
  \mathbf{h}_{k,i};\;
  a_{k,i};\;
  c_{k,i};\;
  w_{k,i}
\Bigr]
\in \mathbb{R}^{D+3}
\label{37}
\end{equation}

5.Prediction heads for anchor interval  offset regression and classification. Firstly, the shared feature \(\mathbf{z}_{k,i}\) is fed into a scale-specific MLP, predicting multiple outputs for each anchor,  the full prediction vector is partitioned as prediction slicing as shown in equation \eqref{38}.

\begin{equation}
\begin{aligned}
  \mathbf{u}^s_{k,i} 
  &= \mathrm{MLP}_k(\mathbf{z}_{k,i})_{1:B_k},
  \quad \mathbf{u}^e_{k,i}
  = \mathrm{MLP}_k(\mathbf{z}_{k,i})_{B_k+1:2B_k} \\
   \mathbf{u}^{\rm conf}_{k,i} 
  &= \mathrm{MLP}_k(\mathbf{z}_{k,i})_{2B_k+1},
  \quad \mathbf{u}^{\rm cls}_{k,i} 
  = \mathrm{MLP}_k(\mathbf{z}_{k,i})_{2B_k+2:2B_k+1+(C-1)}
  \label{38}
\end{aligned}
\end{equation}

where 
$\mathbf{u}^s_{k,i}$,$\mathbf{u}^e_{k,i}$ : represent start and end offset logits into \(B_k\) bins each.

$\mathbf{u}^{\rm conf}_{k,i}$:  Confidence score indicating the quality of the refined interval.

$\mathbf{u}^{\rm cls}_{k,i}$: Class logits predicting event categories (excluding "normal" class).

For the soft offset regression, instead of predicting absolute offset values, we perform distributional regression over discrete bins. The predicted offsets are computed as the expected value over the softmaxed bin logits offset computation in \eqref{39}.

\begin{equation}
\begin{aligned}
\Delta^s_{k,i}
&= \sum_{b=1}^{B_k}
\frac{\exp(u^s_{k,i,b})}
{\sum_{b'}\exp(u^s_{k,i,b'})}
d_{k,b} \\
\Delta^e_{k,i}
&= \sum_{b=1}^{B_k}
\frac{\exp(u^e_{k,i,b})}
{\sum_{b'}\exp(u^e_{k,i,b'})}
d_{k,b}
\label{39}
\end{aligned}
\end{equation}

where \(\{d_{k,b}\}\) are learnable bin centers initialized linearly.

Finally, the refined start and end points of each interval are then given by \eqref{40}:
\begin{equation}
  \text{start}_{k,i} = \alpha_{k,i}L + \Delta^s_{k,i},
  \quad \text{end}_{k,i} = \beta_{k,i}L + \Delta^e_{k,i}
  \label{40}
\end{equation}

These predicted intervals are optionally clamped to \([0, L]\).

\subsection{The proposed  Framework}

In this section, we will introduce  the respiratory sound network architecture and the corresponding loss function design.

\subsubsection{Network Architecture}

As shown in  {Fig.2}, the forward flow of the entire network framework is constructed by integrating four major modules: spectrogram generation, graph construction, node update, and anchor interval  refine. Now, let's walk through each of these modules in turn.

\begin{figure*}[htbp]
\begin{center}
\includegraphics[width=0.85\textwidth]{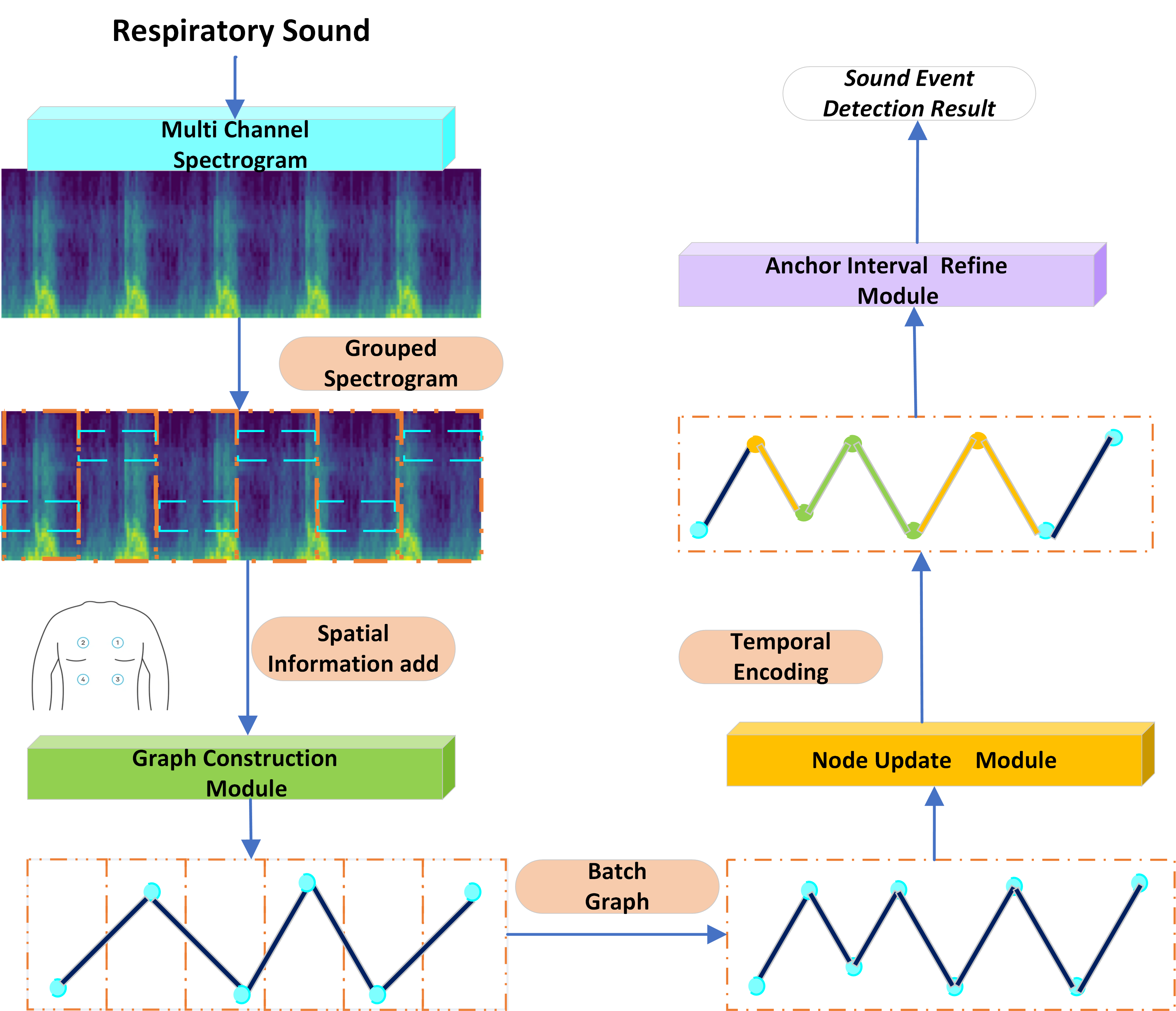}
\end{center}
\caption{The proposed framework based on Graph Neural Network and anchor interval for Respiratory Sound Event Detection}
\label{Fig.2}
\end{figure*}

Firstly, in the spectrogram generation step, we concatenate three types of spectrograms, Mel, gammatone, and CQT  along the dimensional axis to form a multi channel spectrogram feature. Then, for the construction of the graph from the respiratory sound spectrogram, we group the spectrogram along the time-frame dimension, with each group consisting of five frames. Dynamic convolution is used to generate a node feature for each group, optionally incorporating gender and location information to enhance the respiratory sound data. For node labeling, the confidence  reflects the proportion of anomalous frames within each group, while the category label indicates the type of abnormal respiratory sound. Edges between nodes are connected sequentially based on time, and edge attributes are generated through linear mapping of the features of the connected nodes. The edge labels are determined based on the types of the connected nodes. This results in the graph data structure corresponding to each respiratory sound spectrogram. Subsequently, using the edge index and batch attributes, a batch of respiratory sound graphs is constructed.

Afterwards, in the node feature update process, we apply two GAT layers sequentially. Each layer aggregates information from temporally adjacent nodes and updates each node's feature by weighting it according to an attention mechanism based on edge attributes. During the update of each node, the logit values for each node's category are also obtained.

Finally, in the anchor interval refine module, we first introduce cosine temporal encoding to the updated node features and perform Gaussian balancing on the node category logit values. For each anchor interval representing a temporal range, we identify the nodes corresponding to that interval. The node features and node category logit values for each anchor interval are then input into two separate GRUs, generating local node features and local node anomaly scores. These local features, local anomaly scores, and the geometric information of the anchor intervals are aggregated to form a shared feature. This shared feature is used to generate the offset, confidence, and category for each interval. After applying the offset to each corresponding anchor interval, we obtain the final fine-tuned anchor intervals.

\subsubsection{Loss Function Design}

The training  progress  computes a composite loss comprising five distinct components to optimize the GraphRespiratory model for respiratory sound analysis. These losses target both node-level and interval-level predictions, ensuring accurate detection and classification of respiratory events. Each loss is weighted and summed to form the total loss, which guides model optimization. Below, we detail the principle, steps, and mathematical formulation for each loss.

1. Node Confidence Loss: 
The node confidence loss is a regression loss that aligns the predicted confidence logits with ground-truth confidence scores derived from frame-level annotations. This encourages the model to predict the confidence of nodes being abnormal, reflecting the proportion of abnormal frames within each node's temporal span.

\begin{equation}
\begin{aligned}
\mathcal{L}_{\text{node\_conf}} = -\frac{1}{N} \sum_{i=1}^{N} \bigl( t_i \log(\sigma(p_i)) + (1 - t_i) \log(1 - \sigma(p_i)) \bigr)
\label{41}
\end{aligned}
 \end{equation}

 where  the

- \( N \): Total number of nodes in the batch.

- \( p_i \): Predicted confidence logit for node \( i \).

- \( t_i \): Ground-truth confidence score for node \( i \), in \([0, 1]\).

- \( \sigma \): Sigmoid function, \(\sigma(x) = \frac{1}{1 + e^{-x}}\).
 
In \eqref{41},  we use a binary cross-entropy loss with logits (BCEWithLogitsLoss) to handle the continuous nature of confidence scores (ranging from 0 to 1), it computes the BCE loss between predicted logits and target confidences. By extract node confidence logits from the  node predictions$[:, 0]$  and use ground-truth node confidences  as targets to represent the fraction of abnormal frames in each node.

2. Node Classification Loss: 
This loss applies cross-entropy to classify nodes with abnormal labels (excluding normal nodes, labeled as $-1$. It targets the foreground nodes (those with valid abnormal labels) to predict one of the abnormal classes (e.g., 0 to 3 after mapping). This ensure the accurate classification of nodes into normal or abnormal respiratory sound categories, focusing only on abnormal nodes.

During the process, we first identify foreground nodes(which are abnormal nodes) using a mask, then  extract predictions for abnormal classes  via the  node predictions$[:, 1:5]$  and corresponding labels node type labels. Then, compute cross-entropy loss if foreground nodes exist as shown in equation \eqref{42} , otherwise, assign a zero loss.

\begin{equation}
\begin{aligned}
  \mathcal{L}_{\mathrm{node\_cls}} = -\frac{1}{N_{\mathrm{fg}}} \sum_{i \in \mathrm{fg}} \log \left( \frac{\exp(p_{i, y_i})}{\sum_{c=0}^{C-1} \exp(p_{i, c})} \right)
  \label{42}
\end{aligned}
  \end{equation}

where the:

  - \( N_{\text{fg}} \): Number of foreground nodes (where \( y_i \neq -1 \)).

  - \( \text{fg} \): Set of foreground node indices.

  - \( p_{i, c} \): Logit for class \( c \) for node \( i \).

  - \( y_i \): Ground-truth abnormal class label for node \( i \), in \(\{0, 1, 2, 3\}\).

  - \( C \): Number of abnormal classes (e.g., 4).

\textbf{Note}: If $N_{\text{fg}} = 0$, then $\mathcal{L}_{\text{node\_cls}} = 0$.

3.For the Interval Confidence Loss, we use this loss to optimizes the model's ability to predict whether an interval contains an abnormal respiratory event. In  \eqref{43}, it uses BCEWithLogitsLoss to match predicted interval confidence logits with ground-truth confidence labels.

During the process,  we concatenate predicted interval confidence logits across the batch as the batch predicted  confidence, then  concatenate ground-truth confidence labels as the  batch ground truth confidences. It applies to all predicted intervals, ensuring robust event detection.

\begin{equation}
  \begin{aligned}
   & \mathcal{L}_{\text{interval\_conf}} \\
   & = -\frac{1}{M} \sum_{i=1}^{M} \left\{ t_i \log\left(\sigma(p_i)\right) + \left(1 - t_i\right) \log\left(1 - \sigma(p_i)\right) \right\}
    \label{43}
  \end{aligned}
\end{equation}

where the 

  - \( M \): Total number of predicted intervals across the batch.

  - \( p_i \): Predicted confidence logit for interval \( i \).

  - \( t_i \): Ground-truth confidence label for interval \( i \), in \(\{0, 1\}\).

4. Interval Classification Loss,in \eqref{44} this loss is used to classifies intervals identified as containing abnormal events into specific respiratory sound categories. Similar to node classification, this cross-entropy loss targets foreground intervals (those with confidence \( t_i > 0 \)) to predict abnormal class labels. It ensures accurate categorization of detected events.

\begin{equation}
    \mathcal{L}_{\mathrm{interval\_cls}} = -\frac{1}{M_{\mathrm{fg}}} \sum_{i \in \mathrm{fg}} \log \left( \frac{\exp(p_{i, y_i})}{\sum_{c=0}^{C-1} \exp(p_{i, c})} \right)
    \label{44}
\end{equation}

where the 

  - \( M_{\text{fg}} \): Number of foreground intervals (where \( t_i > 0 \)).

  - \( \text{fg} \): Set of foreground interval indices.

  - \( p_{i, c} \): Logit for class \( c \) for interval \( i \).

  - \( y_i \): Ground-truth abnormal class label for interval \( i \), in \(\{0, 1, 2, 3\}\).

  - \( C \): Number of abnormal classes (e.g., 4).

  \textbf{Note}: If $M_{\text{fg}} = 0$, $\mathcal{L}_{\text{interval\_cls}} = 0$.

In this process, we apply a foreground mask to select intervals with events, then, extract classification logits and corresponding labels. After that, compute the cross-entropy loss if foreground intervals exist, otherwise, assign zero loss.

5. In the Interval Localization Loss \eqref{45} ,  we use a negative log Intersection over Union (IoU) metric to measure the overlap between predicted and ground-truth interval boundaries. It focuses on to refine the temporal boundaries of predicted intervals to align with ground-truth event locations.

During this process, we select predicted boundaries and ground-truth boundaries for foreground intervals. Then, compute the negative log IoU loss, which penalizes deviations in start and end times. If there is  no foreground intervals exist we assign the zero loss.

\begin{equation}
  \begin{aligned}
    \mathcal{L}_{\text{interval\_loc}} = -\frac{1}{M_{\text{fg}}} \sum_{i \in \text{fg}} \log \left( \mathrm{IoU}(B_i^{\text{pred}}, B_i^{\text{gt}}) \right)
    \label{45}
  \end{aligned}
\end{equation}

where the 

  - \( M_{\text{fg}} \): Number of foreground intervals.

  - \( \text{fg} \): Set of foreground interval indices.

  - \( B_i^{\text{pred}} = [s_i^{\text{pred}}, e_i^{\text{pred}}] \): Predicted interval boundaries (start, end) for interval \( i \).

  - \( B_i^{\text{gt}} = [s_i^{\text{gt}}, e_i^{\text{gt}}] \): Ground-truth interval boundaries.

  - \( \text{IoU}(B_i^{\text{pred}}, B_i^{\text{gt}}) = \frac{\min(e_i^{\text{pred}}, e_i^{\text{gt}}) - \max(s_i^{\text{pred}}, s_i^{\text{gt}})}{\max(e_i^{\text{pred}}, e_i^{\text{gt}}) - \min(s_i^{\text{pred}}, s_i^{\text{gt}})} \), clipped to \([0, 1]\).

  \textbf{Note}: If $M_{\text{fg}} = 0$, then $\mathcal{L}_{\text{interval\_loc}} = 0$.

6.Finally, for the Total Loss we combine the all five components, as  shown in \eqref{46} each scaled by its respective weight to balance their contributions.

During the training process, each loss is logged individually and the total loss  is tracked for monitoring training progress.

\begin{equation}
  \begin{aligned}
    \mathcal{L}_{total} 
    & = w_{\text{n\_conf}} \mathcal{L}_{\text{node\_conf}} + w_{\text{n\_cls}} \mathcal{L}_{\text{node\_cls}}  \\
    & + w_{\text{i\_conf}} \mathcal{L}_{\text{interval\_conf}} 
      + w_{\text{i\_cls}} \mathcal{L}_{\text{interval\_cls}} \\
    & + w_{\text{i\_loc}} \mathcal{L}_{\text{interval\_loc}}
    \label{46}
  \end{aligned}
\end{equation}

where the weights are defined as follows:

  - \( w_{\text{n\_conf}} \): node confidence loss weight.

  - \( w_{\text{n\_cls}} \): node classification loss weight.

  - \( w_{\text{i\_conf}} \): interval confidence loss weight.

  - \( w_{\text{i\_cls}} \): interval classification loss weight.

  - \( w_{\text{i\_loc}} \): interval localization loss weight.

\section{Experiments}
In the experimental part, we  introduce the data set and data preprocessing, environment and experimental settings, as well as experimental results and analysis one by one.

\subsection{Dataset and Data Preprocessing}

\subsubsection{Dataset}

This study utilizes the SPRSound database \cite{zhang2024meta}, which contains pediatric respiratory sounds sampled at 8 kHz with a 16-bit resolution. Pediatricians annotated these recordings with various sound events, establishing a gold-standard reference for the dataset. The respiratory sounds are categorized into normal, rhonchi, wheeze, stridor, coarse crackle, fine crackle, and wheeze $\&$ crackle. SPRSound provides labeled annotations for each event, enabling the study of event detection and classification.

In order to verify the generalization of the model, we also verified our mothod on the HF lung v1 \cite{hsu2021benchmarking} dataset, which is a dataset with a sampling rate of 4 kHz and collected using a stethoscope 3M.

\subsubsection{ Data Preprocessing}

During preprocessing, for the SPRSound database   all audio is sampled to 8 kHz and keep the original length of each audio. Parameters used in the Short-Time Fourier Transform (STFT) for converting raw audio to spectrograms are set as follows: the number of FFT points $n\_fft$ = 1024, window length in points $win\_len$ = 1000, hop length in points $hop\_len$ = 128. The lowest and highest cutoff frequencies in spectrogram features are set to $f\_min$ = 32.7 and $f\_max$ = 4000 respectively, with a uniform filter bank size of 84.

\subsubsection{Data Augmentation}

To address data imbalance, data augmentation is performed at both the raw audio and spectrogram levels. During preprocessing, each batch of audio samples is randomly augmented using techniques such as adding noise,  time-stretching, and  Vocal Tract Length Perturbation (VTLP) .

In particular, for the sound event detection task, we applied time shift data enhancement, which will move the area where the abnormal event occurs forward or backward. It should be noted here that if an event occurs at the beginning and end of the audio during the movement, then it is very likely that after applying time shift, the event will be split into the beginning and the end, resulting in an increase in the number of abnormal events in the audio. Moreover, after we apply time shift, the real label corresponding to our anchor interval also needs to be updated accordingly.

During training, the generated spectrograms are further augmented by applying random masking along either the frequency or time dimensions.

\subsubsection{Spectrogram Normalization}

Unlike conventional methods that normalize the entire spectrogram, this study normalizes each row of the spectrogram independently. Each row corresponds to a specific frequency band, while different rows represent different frequency ranges. Row-wise normalization enhances the detection of variations within a single frequency band. Conversely, each column represents information across frequencies, where most respiratory sound features are concentrated in the low-frequency range. Column-wise normalization risks suppressing high-frequency anomalies due to the dominance of low-frequency components. Therefore, row-wise normalization is adopted to preserve critical respiratory sound features.

\subsubsection{ Frames in each group }

The total number of frames per sample is denoted as $T_{frames}$, and the spectrogram is slid right by $S_{frames}$ frames each group. 
 
The number of groups obtained per sample is given by $N_{g}= \frac {T_{frames}}{ S_{frames}}$. Moreover, the number of frames to slide, $S_{frames}$, is set equal to  the number of frames in each group,  ensuring that there are no overlapping parts between groups, which is beneficial for temporal detection, in our experiment we set  each group has five frames.

\subsection{Experiment}

\subsubsection{ Experiment  Environment }

The experiments were conducted on an  Ubuntu 20.04  system using a Python 3.8 virtual environment created with Anaconda3. The deep learning framework employed is  PyTorch 1.13.1-cuda 11.7  with  PyTorch-Geometric 2.5.2 . The hardware setup includes an  Intel Core i9 processor, 24 GB RAM, and an  NVIDIA GeForce RTX 3090 GPU  with  24 GB memory .

\subsubsection{ Optimizer Configuration}
In the network training process, we use the PyTorch Lightning training loop and employ a dual-optimizer strategy to separately optimize node-related and interval-related parameters, enhancing training stability and flexibility.

For node parameters optimizer, it includes shared parameters  and node confidence/classification head parameters, this component involved in node feature extraction and classification (e.g., CNN, GAT, node classifier).

And in the interval parameters optimizer, it includes interval refinement head parameters, this components responsible for interval localization and classification (e.g., GRU-based refinement layers). Both  optimizers use  the  Adam with an initial learning rate of $1 \times 10^{-3}$.

For the learning rate schedulers, we also  use two different strategies,  in  node parameters scheduler, we use LambdaLR with exponential decay, learning rate decays as $\mathrm{lr} = \mathrm{lr}_{\mathrm{initial}} \times 0.99^{\mathrm{step} / 126}$, mimicking a step-wise decay over 126 batches per epoch and updated every step. The node parameters scheduler gradually reduces the learning rate to stabilize node-related optimization.

In the interval parameters scheduler, we use CosineAnnealingLR with learning rate oscillates over a period of $T_{\mathrm{max}} = 50400$ steps (400 epochs $\times$ 126 batches), optionally with a minimum learning rate (e.g., $2 \times 10^{-4}$) and updated every step. this scheduler provides dynamic learning rate adjustments to improve convergence for interval-related tasks, which require precise boundary predictions.

\subsubsection{Evaluation Metrics for Sound Event Detection}

Given the limited research on respiratory sound event detection, this study employs the fundamental event-based metrics as shown in {Fig.3} for evaluation. These metrics compare system outputs with ground truth to generate performance indicators for event detection and classification. The key evaluation metrics include  F1-score (F) in equation  \eqref{47}  and  Error Rate (ER), with tunable parameters such as time tolerance $onset$  and duration tolerance  $offset$ . In our evaluation, a fixed 200 ms tolerance is applied for onset, while either 200 ms or 10$\%$ of the event's length is used for offset tolerance.

\begin{figure}[htbp]
  \begin{center}
  \includegraphics[width=0.4\textwidth]{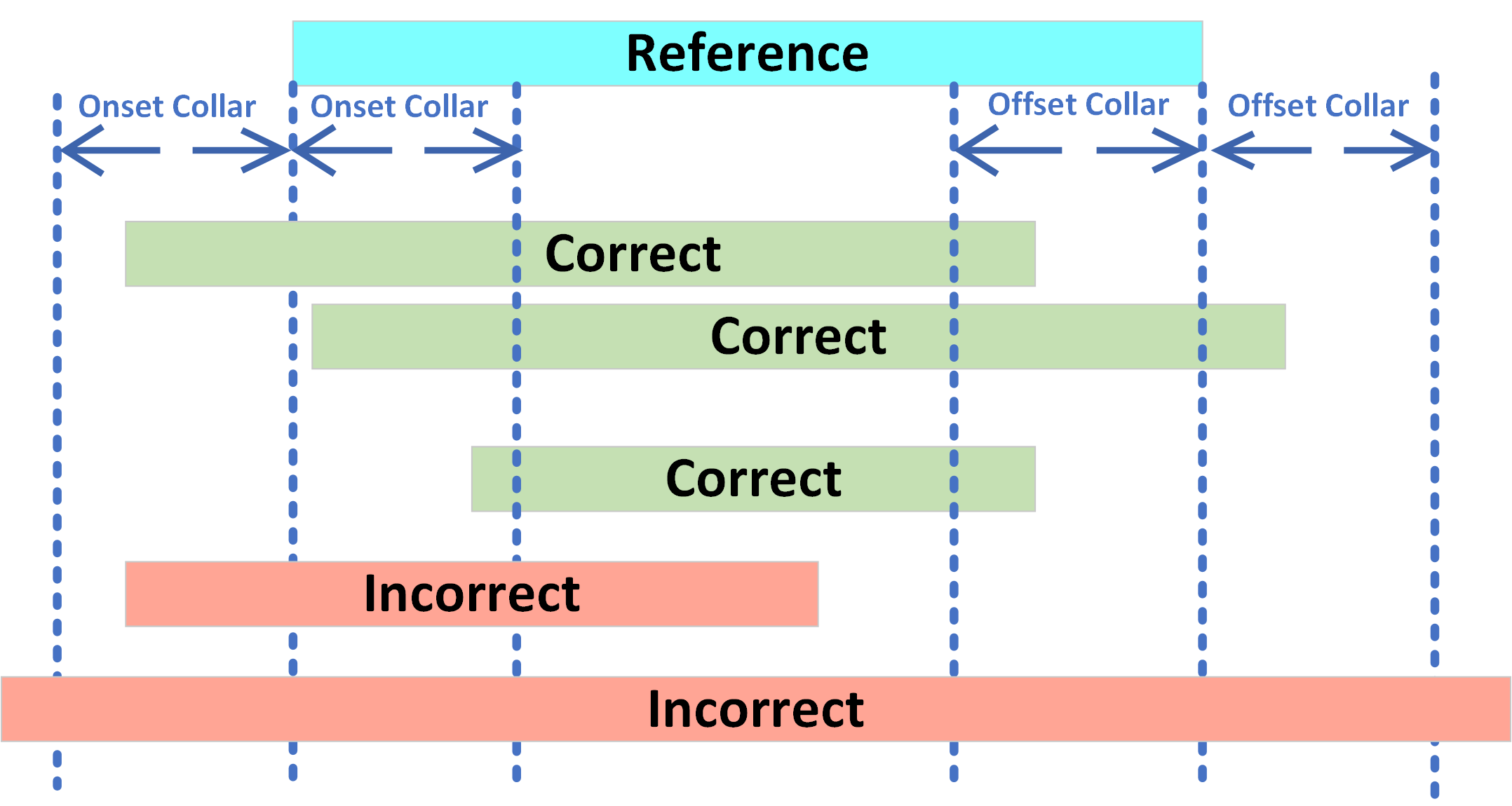}    
  \end{center}
  \caption{Event-based evaluation metrics with fixed 200ms collar on onsets and  200ms/10 $\%$  of the event’s length collar on offsets.}
  \label{Fig.3}
  \end{figure}

\begin{figure}[htbp]
  \begin{center}
  \includegraphics[width=0.4\textwidth]{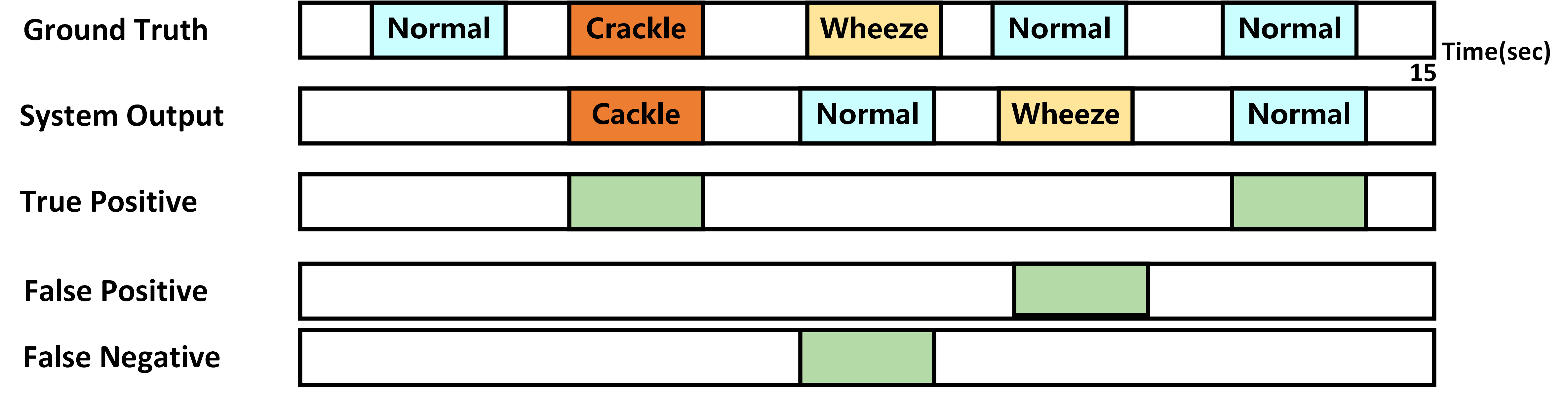}    
  \end{center}
  \caption{  Calculation of the event-based evaluation metrics.}
  \label{Fig.4}
  \end{figure}

{Fig.4} illustrates the evaluation principle using a 15-second audio file containing multiple events. By comparing system outputs with ground truth, we identify  false negatives (FN) ,  false positives (FP) , and  true positives (TP) . An event is considered a true positive if it matches both the timing and event type in the system output and ground truth; otherwise, it is classified as an error. Consequently, the overall metrics are dominated by events with timing or classification errors.

\begin{equation}
  \begin{aligned}
  F1 \; \text{Score}
  & = \frac{2 \cdot TP}{2 \cdot TP + FP + FN}\\
  ER &= \frac{S + D + I}{N} \\ 
  S &= \min(FN, FP) \\
  D &= \max(0, FN - FP) \\
  I &= \max(0, FP - FN) \\
  \label{47}
  \end{aligned}
  \end{equation}

In equation  \eqref{47} ,$N$ is the number of respiratory events  marked as active in the annotation. S is the minimum value between FP and FN. D and I represent deletion errors and insertion errors, respectively.

\begin{itemize}
 \item [1)] 
F1-score (F):  is critical in our evaluation. It measures the accuracy of the system by considering TP, FP, and FN, providing a comprehensive assessment of performance.

\item [2)] 
Error Rate (ER): 
The error rate, measures the proportion of mispredicted events. It also incorporates elements considered in the F-score, with additional variables defined as follows:

\end{itemize}

By analyzing these metrics, the system’s detection timing and classification accuracy can be effectively evaluated.

\subsection{Experiments Result}

\subsubsection{Influence of integrated head and separate head}

In this section of the experiments, we investigated the impact of edge attribute generation methods and interval offsets resolution under two configurations: integrated head and separate head.

Inspired by the design of YOLO in computer vision, the integrated head adopts the idea that the presence of a target, its type, and its location should be tightly coupled. In this approach, a shared feature is used as input to the integrated head, which jointly learns the anchor interval offsets, confidence, and classification. The parameters are then sliced and extracted for separate optimization using different loss functions.

As shown in {Table1}, the performance of edge attribute generation using compressed methods under the integrated head outperforms sequential generation. This is  due to the model's ability to actively filter redundant and irrelevant information during the compression process, thus improving the distinction between edges.

\begin{table}[ht]
  \centering
  \caption{Detection performance under the integrated head, where S represents sequence and C represents compression.} 
  \begin{tabular}{lcccc}
  \toprule
  \textbf{Edge Att} & \textbf{Offset Res} & \textbf{Step } & \textbf{F1 (\%)}  & \textbf{Interval Cls loss }\\ 
  \midrule
   C          & [-0.5, 0.5]           & 16k          & 14.8  & 0.47    \\ 
   C          & [-1.0, 1.0]           & 15k          & 19.7   & 0.60    \\ 
   C          & [-1.5, 1.5]           & 20k          & 16.2   & 0.46    \\ 
   C          & [-20.0, 20.0]         & 22k          & 22.3   & 0.54    \\ 
   S          & [-0.5, 0.5]           & 20k          & 9.4    & 0.62    \\ 
   S          & [-1.0, 1.0]           & 20k          & 21.1   & 0.44    \\ 
   S          & [-1.5, 1.5]           & 20k          & 7.5   & 0.51    \\
   S          & [-20.0, 20.0]         & 20k          & 17.9   & 0.50    \\ 
  \bottomrule
  \end{tabular}
  \label{Table1}
  \end{table}

Although the integrated head is computationally efficient, it has a significant proportion of shared parameters for the anchor interval offsets, with much fewer parameters dedicated to confidence and classification. To investigate whether learning the interval offsets suppresses the learning of confidence and classification, we designed the separate head method, decoupling the interval offsets from confidence and classification, with distinct heads for each.

  \begin{table}[ht]
    \centering
    \caption{Detection performance under the separation head, where S represents sequence and C represents compression.} 
    \begin{tabular}{lcccc}
    \toprule
    \textbf{Edge Att} & \textbf{Offset Res} & \textbf{Step } & \textbf{F1 (\%)}  & \textbf{Interval Cls loss }\\ 
    \midrule
     C          & [-0.2, 0.2]           &  22k         &  11.8  &  0.54   \\
     C          & [-0.5, 0.5]           &  15k         &  22.1  &  0.44   \\ 
     C          & [-1.0, 1.0]           &  20k         &  10.3  &  0.62   \\
    C           & [-1.5, 1.5]         & 20k          & 12.6   & 0.45   \\
     C          & [-10.0, 10.0]       & 20k          & 8.5   & 0.54 \\   
     C          & [-20.0, 20.0]         &  20k         &  4.5   &  0.45  \\ 
     S          & [-0.5, 0.5]           & 19k          & 11.2   & 0.45    \\ 
     S          & [-1.0, 1.0]           & 18k          & 12.2   & 0.51    \\
     S          & [-1.5, 1.5]           & 20k          & 19.1   & 0.53    \\ 
     S          & [-20.0, 20.0]         & 16k          & 12.2   & 0.61    \\ 
    \bottomrule
    \end{tabular}
    \label{Table2}
    \end{table}

{Table2} shows that, under the separate head configuration, the performance of edge attribute generation via compressed methods remains superior to sequential generation. However, in the separate head design, overall performance is significantly affected by the anchor interval offset displacement.

\begin{figure*}[htbp]
\centering
\includegraphics[width=0.95\textwidth]{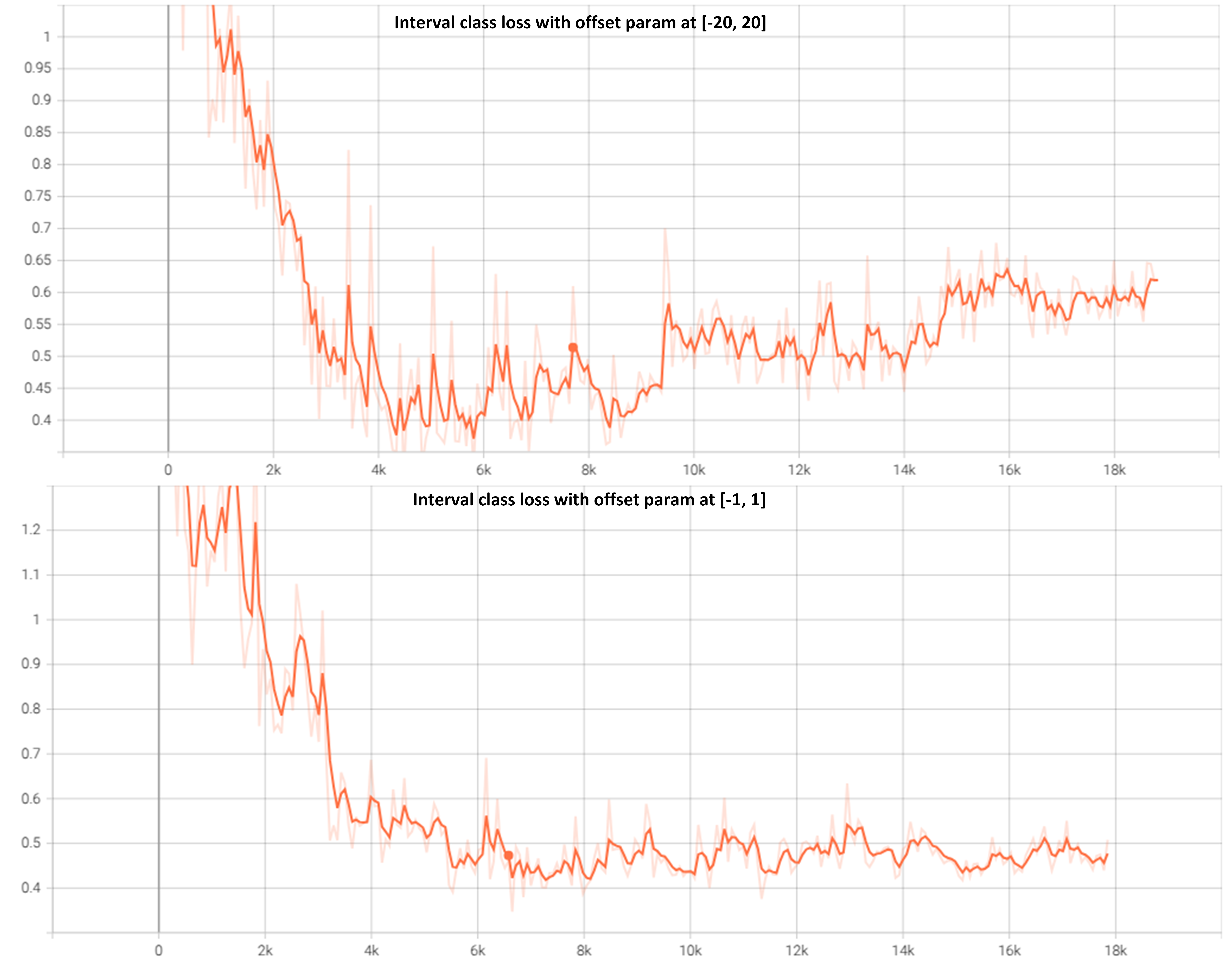}    
\caption{The convergence of interval classification losses under different offset configurations}
\label{Fig.5}
\end{figure*}

 As shown in {Fig.5}, the range of interval  displacement affects the stability of model convergence during training. The top and bottom  of {Fig.5} display the model's convergence on event interval classification loss under interval displacement configurations of [-20, 20] and [-1, 1], respectively.

Overall, the integrated head approach outperforms the separate head configuration, and the compressed generation of edge attributes yields better results than sequential generation. Moreover, the integrated head demonstrates stronger robustness to interval displacement compared to the separate head.

\subsubsection{ Comparison and Ablation  stduy }

In this section, we compare our work with existing methods in the abnormal respiratory sound event detection task and conduct ablation experiments on the incorporation of respiratory sound position information.

\begin{table*}[ht]
  \centering
  \caption{Comparing with other method on Respiratory Sound Event Detection}
  \begin{tabular}{lccccc} 
  \toprule
  \textbf{Method} & \textbf{Dataset}  & \textbf{Class wise F1 (\%)}  & \textbf{ER} & \textbf{Model size}  \\ 
  \midrule
  CNN+VAD \cite{yeh2024novel}         & SPR Sound & 6.81 & 1.26  &   65MB   \\ 
  CNN+BiLSTM \cite{hsu2021benchmarking} & HF Lung & - & -    & 280MB+  \\ 
  Ours                                & SPR Sound & 22.3 & 1.08   & 99.7MB       \\ 
  Ours                                &  HF Lung & 16.8 & 1.55      & 99.7MB    \\  
  \bottomrule
  \end{tabular}
  \label{Table3}
  \end{table*}

As shown in {Table3}, on the SPR Sound dataset, we compare our method with the work of \cite{yeh2024novel} et al. Both approaches perform event detection with four types of abnormal respiratory sounds. The key difference lies in their consideration of normal respiratory sounds, \cite{yeh2024novel}  et al. also include normal sounds in the detection process, while our experiments exclude them. We consider that, in the context of abnormal respiratory sound event detection, only the detection of potential abnormal sounds is necessary, and intervals not classified as abnormal can be assumed to correspond to normal respiratory sounds.

On the HF Lung v1 dataset, we compare our method with the work of \cite{hsu2021benchmarking} et al. However, since their paper does not provide class-wise level detection performance for abnormal respiratory events, and they do not perform detection for all four abnormal sound types simultaneously but rather separate continuous and discontinuous abnormal sounds (e.g., wheeze and crackle), we cannot directly compare their performance under identical evaluation settings. Given the differences in experimental setup and evaluation metrics, the results presented here only include the model weight size used by \cite{hsu2021benchmarking} et al.

Moreover, the method used by\cite{yeh2024novel}  et al. involves a two-stage process: first, a classification model is trained using respiratory sound spectrograms, followed by the detection of abnormal sound time intervals using a separate sound event detection model. The detected intervals are then converted into spectrogram features, which are classified using the initial trained model. This separation of interval region detection and event classification is not an end-to-end learnable framework. In contrast, the approach of \cite{hsu2021benchmarking}  uses CNN$+$BiLSTM or CNN+BiGRU, which are end-to-end learnable architectures that are feasible on server-level machines. However, due to the large parameter count and computational demands, they face challenges related to memory, processing power, and energy consumption, hindering their deployment on mobile and edge devices.

In comparison, the architecture proposed in this work achieves end-to-end learnable event detection for all four abnormal respiratory sounds, while maintaining reasonable model weight sizes. Given that abnormal respiratory sound event detection is a relatively new research area with limited studies, the existing methods still show considerable space for improvement, as current performance does not yet meet satisfactory standards.

\begin{table}[ht]
\centering
\caption{Node-level classification performance without respiratory position information}
\begin{tabular}{lccc}
\toprule
\textbf{Category} & \textbf{Precision}  & \textbf{Recall} & \textbf{F1 Score} \\
\midrule

Rhonchi &  0.4276  & 0.4256  & 0.4266 \\
Wheeze  &  0.6185  & 0.3804  & 0.4711 \\

Stridor & 0.2614    & 0.2879  & 0.2740 \\
Crackle & 0.7322    & 0.8590  & 0.7905 \\

\bottomrule
\end{tabular}
\label{Table4}
\end{table}

\begin{table}[ht]
\centering
\caption{Node-level classification performance with respiratory position information}
\begin{tabular}{lccc}
\toprule
\textbf{Category} & \textbf{Precision}  & \textbf{Recall} & \textbf{F1 Score} \\
\midrule

Rhonchi &  0.4199    & 0.4456 & 0.4325 \\
Wheeze  &  0.5759    & 0.4204 & 0.5011 \\

Stridor & 0.2460    & 0.3379 & 0.3040 \\
Crackle & 0.7252    & 0.8621 & 0.7804  \\

\bottomrule
\end{tabular}
\label{Table5}
\end{table}

On the other hand,   {Table4} and {Table5} show the model’s performance at the node level with and without respiratory sound positional information. The results demonstrate that incorporating positional information improves the recall rates for all four abnormal respiratory sounds, indicating that respiratory position information helps the system better recognize abnormal respiratory sounds.

\subsubsection{Generalization experiment }

To evaluate the generalization capability of the proposed method, experiments were conducted on two datasets: SPR \cite{zhang2024meta} Sound 2024 and HF Lung v1 \cite{hsu2021benchmarking}.

\begin{figure*}[htbp]
\begin{center}
\includegraphics[width=0.95\textwidth]{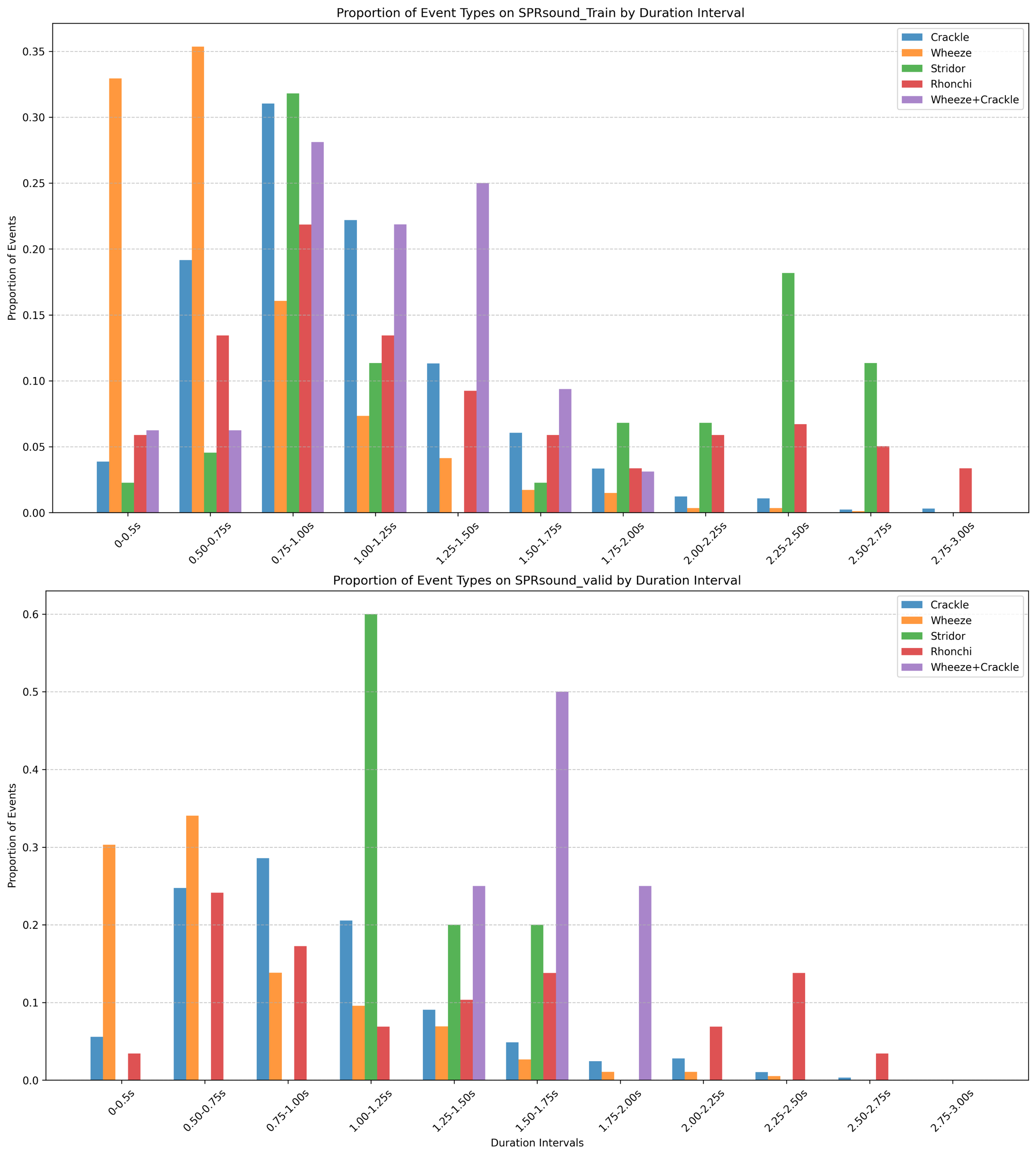}    
\end{center}
\caption{In the SPR Sound dataset, the distribution ratio of different abnormal respiratory sounds in various duration periods. The upper one is the training set, and the lower one is the validation set.}
\label{Fig.6}
\end{figure*}

\begin{figure*}[htbp]
  \begin{center}
  \includegraphics[width=0.95\textwidth]{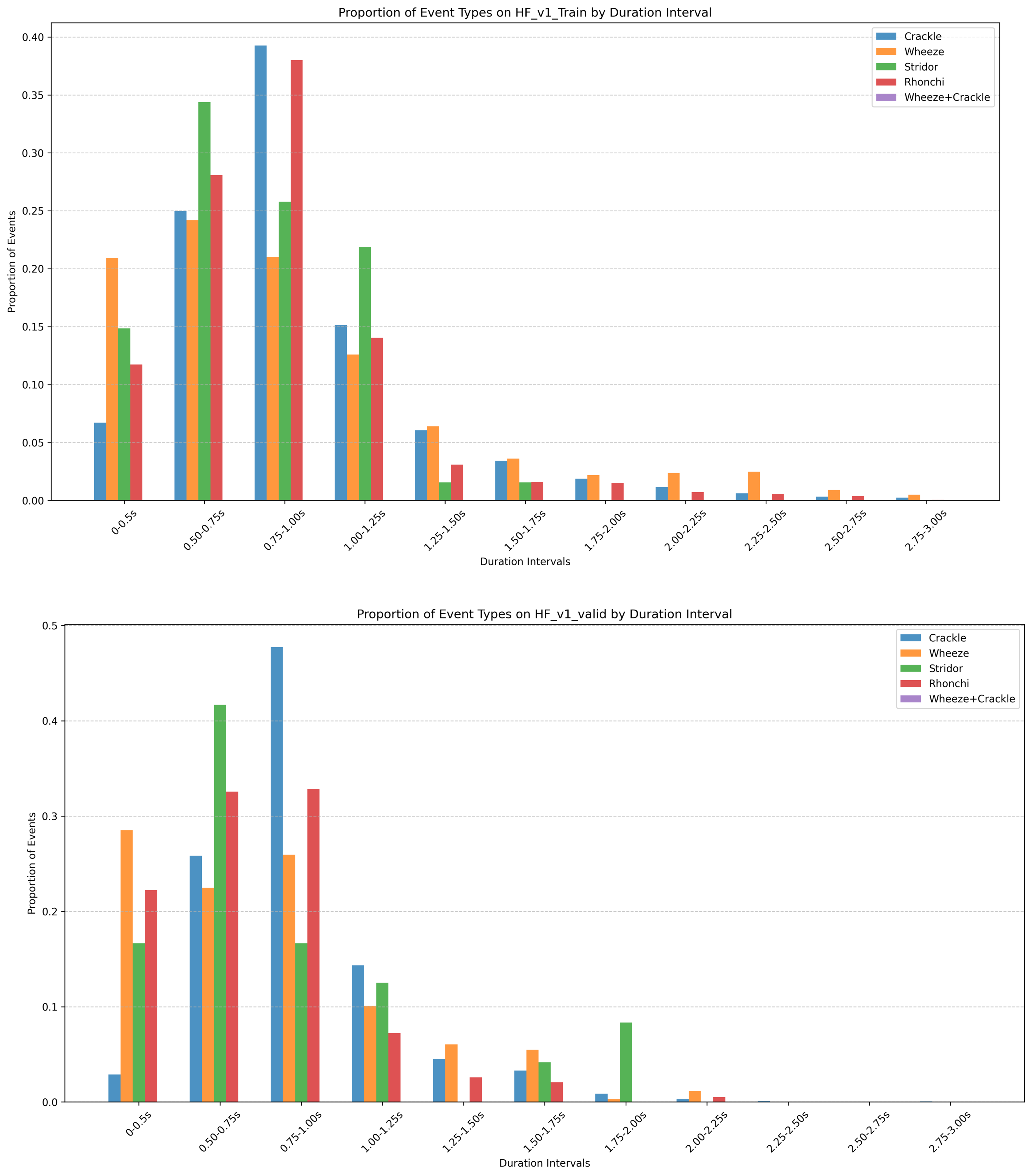}    
  \end{center}
  \caption{In the HF Lung v1 dataset, the distribution ratio of different abnormal respiratory sounds in various duration periods. The upper one is the training set, and the lower one is the validation set.}
  \label{Fig.7}
  \end{figure*}

{Fig.6} and {Fig.7} present the distribution of four abnormal respiratory sounds (crackle, wheeze, stridor, and rhonchi) in both datasets. The duration of these sounds was observed at four intervals: 0.5s, 0.75s, 1s, and 1.25s, which align with the designed prior anchor intervals of 0.5s, 0.75s, and 1.5s.

 {Table6} and {Table7} show the class-wise evaluation metrics on the SPR Sound 2024 and HF Lung v1 datasets, respectively, reflecting the precision of event detection for each type of abnormal sound.

\begin{table*}[ht]
\centering
\caption{Respiratory sound event detection results on the SPR sound dataset}
\begin{tabular}{lcccccccc} 
\toprule
\textbf{Event Label} & \textbf{Nref} & \textbf{Nsys} & \textbf{F (\%)} & \textbf{Pre (\%)} & \textbf{Rec (\%)} & \textbf{ER} & \textbf{Del} & \textbf{Ins} \\ 
\midrule
Wheeze   & 188 & 56  & 20.5  &44.6  &13.3    &1.03   &0.87   &0.16     \\ 
Rhonchi  & 29  & 13   &19.0  &30.8  &13.8    &1.17   &0.86  & 0.31   \\ 
Stridor  & 5   & 1   & 33.3  &100.0  &20.0    & 0.80   &0.80   &0.00    \\ 
Crackle  & 287 & 178   & 16.3  &21.3  &13.2   & 1.36  & 0.87  & 0.49       \\ 
\bottomrule
\label{Table5}
\end{tabular}
\end{table*}

\begin{table*}[ht]
  \centering
  \caption{Respiratory sound event detection results on the HF lung v1 dataset}
  \begin{tabular}{lcccccccc} 
  \toprule
  \textbf{Event Label} & \textbf{Nref} & \textbf{Nsys} & \textbf{F (\%)} & \textbf{Pre (\%)} & \textbf{Rec (\%)} & \textbf{ER} & \textbf{Del} & \textbf{Ins} \\ 
  \midrule
  Wheeze   & 1430    &542    &11.8  &21.4  &8.1    &1.22  & 0.92   &0.30      \\ 
  Rhonchi  & 960     &369    &14.7  &26.6  &10.2   &1.18   &0.90   &0.28      \\ 
  Stridor  &29      &30     &3.4   &3.3   &3.4     &1.97   &0.97   &1.00   \\ 
  Crackle  & 1812   &3054   &32.0  &25.5  &43.0    &1.83  & 0.57  & 1.26   \\ 
  \bottomrule
  \label{Table6}
  \end{tabular}
  \end{table*}

From the {Table6} and {Table7}, we can see that for the wheeze and rhonchi events, the system's output was generally lower than the actual number of occurrences in both datasets. Furthermore, the recall rates for continuous abnormal sounds, such as wheeze, rhonchi, and stridor, were notably low. In contrast, the recall rate for crackle, a discontinuous sound, was significantly higher in the HF Lung v1 dataset. This discrepancy is attributed to the feature extraction method, where we used a grouping interval of five frames. This scale effectively captures discontinuous respiratory sounds, but it is less suited for continuous respiratory sounds. Therefore, increasing the number of frames per group is necessary for future improvements in detecting continuous abnormal sounds.

Regarding the detection of stridor, the system produced fewer outputs in the SPR Sound dataset, primarily due to the significant distribution mismatch of stridor in the training and validation sets. However, in the HF Lung v1 dataset,  there was no such mismatch, and the output number was consistent with the actual occurrences. This suggests that the prior anchor intervals at different scales were well-matched to the stridor event in the HF Lung v1 dataset.

For crackle detection, the system’s output was lower in the SPR Sound dataset compared to the HF Lung v1 dataset. This difference is due to the dynamic nature of audio sample lengths in the SPR Sound dataset (ranging from 9.2s to 15.3s), which introduces instability during training. In contrast, the HF Lung v1 dataset standardized all audio samples to 15s, reducing this fluctuation. As illustrated in {Fig.8}, this length variability contributed to jitter in the model’s training process.

\begin{figure*}[htbp]
\begin{center}
\includegraphics[width=0.95\textwidth]{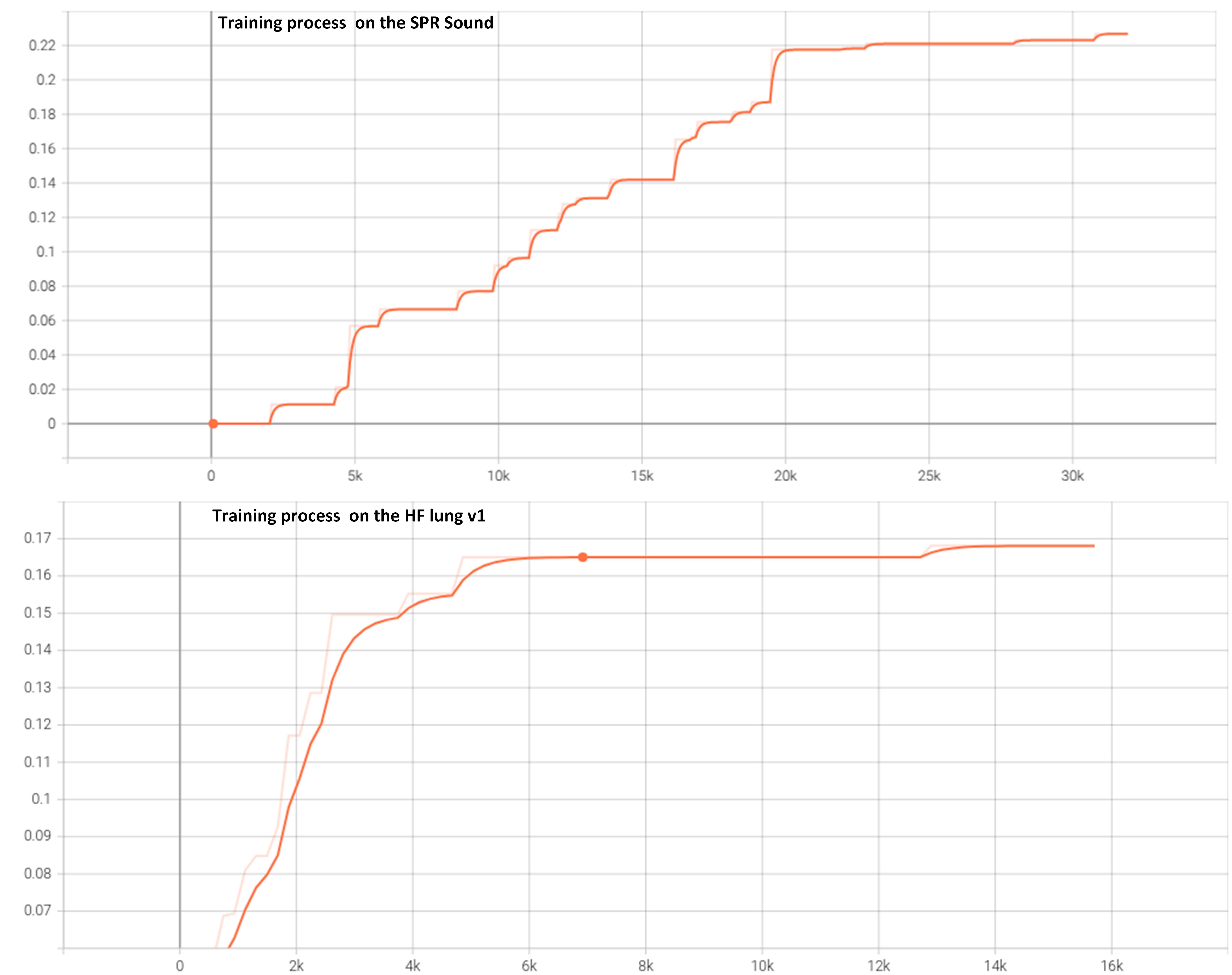}    
\end{center}
\caption{The training process of the proposed method on different datasets. The top one is on the SPR sound dataset, and the bottom one is on the HF lung v1 dataset.}
\label{Fig.8}
\end{figure*}

\subsection{Discussions and Feature work}

Although we have implemented an end-to-end learnable respiratory sound detection framework, there are still areas for improvement in our work:

\textbf{Anchor Interval Fixed Size}: Currently, we use fixed-size anchor intervals to detect potential abnormal segments within an audio clip. This essentially represents an exhaustive approach, covering only three scales. However, as the number of scales increases, while this may improve the model's performance, it also raises the computational cost and training time.

To address this, we plan to explore the use of scalable intervals or allow the model to adaptively learn the optimal interval scale. This will improve the model's generalization while minimizing the computational burden, especially when detecting events in datasets with significant distribution differences.

\textbf{Feature-based Pre-selection}: In the process of feature-based recognition, our current work does not implement a feature pre-selection mechanism. In contrast, works such as \cite{yeh2024novel}, \cite{kao2018r}, and \cite{chaudhary2021automatic} have followed a similar research approach, where after transforming the spectrogram features, they do not directly perform recognition based on these features. Instead, they first select potential anomalous features and then perform anomaly detection based on the filtered features.

This pre-selection mechanism essentially helps the model focus more on anomalous features, improving task performance. This has inspired us to introduce a similar focus mechanism in future work. Specifically, we aim to enhance the model’s ability to distinguish between node’s edge in the graph, enabling it to learn to focus on relevant regions of the audio and thereby improve abnormal feature detection.

\textbf{Training Time and Variable-Length Audio}: Although our graph neural network framework supports the processing of variable-length audio, due to the current implementation of PyTorch Lightning and the Dataloader's $collate \;fn$ function, we were unable to batch process variable-length inputs. As a result, we resorted to a sequential loading approach. This approach incurs significant communication time between the CPU and GPU, which increases the overall training time. Unfortunately, we have not yet found a solution to reduce the training time caused by this issue. We hope to address this in future work and explore potential optimizations.

\section{Conclusions }
In this paper, we introduce graph neural network (GNN) techniques and anchor-based concepts from computer vision object detection into the task of respiratory sound event detection. We design an end-to-end, learnable framework based on GNNs and anchor intervals for sound event detection. Specifically, we first perform grouped feature extraction on the spectrogram and construct a graph data structure for respiratory sounds, enabling the processing of variable-length audio inputs through GNNs. Secondly, inspired by object detection in computer vision, we propose the use of anchor intervals in sound event detection. By decomposing the abnormal respiratory sound event detection into three sub-tasks, interval confidence prediction,  classification,  and interval location regression, we avoid the traditional frame-based post-processing approach, allowing the network to directly learn the event interval boundaries. This results in an end-to-end framework for detecting abnormal respiratory sound events.

Experiments on the SPRSound and HF Lung V1 datasets demonstrate the effectiveness of the proposed method. Ablation studies further indicate that incorporating respiratory sound position information enhances the system’s ability to distinguish between different abnormal respiratory sounds.

\printcredits

\section{Declaration of Competing Interest }
The authors declare that they have no  known competing financial interests or personal relationships that could have appeared to influence the work reported in this paper.

\section{Acknowledgment}
This work was supported in part by the National Natural Science Foundation of China under Grant 2022YFC2404401 and the Hainan Provincial Natural Science Foundation of China under Grant .

\bibliographystyle{ieeetr}
\bibliography{cas-refs}

\begin{thebibliography}{1}

\bibitem{lang2020graph}
R.~Lang, R.~Lu, C.~Zhao, H.~Qin, and G.~Liu, ``Graph-based semi-supervised one class support vector machine for detecting abnormal lung sounds,'' {\em Applied Mathematics and Computation}, vol.~364, p.~124487, 2020.

\bibitem{lang2021analysis}
R.~Lang, Y.~Fan, G.~Liu, and G.~Liu, ``Analysis of unlabeled lung sound samples using semi-supervised convolutional neural networks,'' {\em Applied Mathematics and Computation}, vol.~411, p.~126511, 2021.

\bibitem{jacome2019convolutional}
C.~J{\'a}come, J.~Ravn, E.~Holsb{\o}, J.~C. Aviles-Solis, H.~Melbye, and L.~Ailo~Bongo, ``Convolutional neural network for breathing phase detection in lung sounds,'' {\em Sensors}, vol.~19, no.~8, p.~1798, 2019.

\bibitem{hsu2021benchmarking}
F.-S. Hsu, S.-R. Huang, C.-W. Huang, C.-J. Huang, Y.-R. Cheng, C.-C. Chen, J.~Hsiao, C.-W. Chen, L.-C. Chen, Y.-C. Lai, {\em et~al.}, ``Benchmarking of eight recurrent neural network variants for breath phase and adventitious sound detection on a self-developed open-access lung sound database—hf\_lung\_v1,'' {\em PLoS One}, vol.~16, no.~7, p.~e0254134, 2021.

\bibitem{hsu2023dual}
F.-S. Hsu, S.-R. Huang, C.-F. Su, C.-W. Huang, Y.-R. Cheng, C.-C. Chen, C.-Y. Wu, C.-W. Chen, Y.-C. Lai, T.-W. Cheng, {\em et~al.}, ``A dual-purpose deep learning model for auscultated lung and tracheal sound analysis based on mixed set training,'' {\em Biomedical Signal Processing and Control}, vol.~86, p.~105222, 2023.

\bibitem{yeh2024novel}
C.-Y. Yeh, S.-A. Chiu, X.-Y. Deng, and W.-C. Fang, ``A novel ai-inspired method and system implementation for detecting and classifying pediatric respiratory sound events,'' in {\em 2024 IEEE Biomedical Circuits and Systems Conference (BioCAS)}, pp.~1--5, IEEE, 2024.

\bibitem{zhang2024meta}
Q.~Zhang, C.~Chen, S.~Yuan, J.~Zhang, J.~Yuan, H.~Huang, Y.~Zhang, R.~Pan, X.~Jiang, J.~Zhao, {\em et~al.}, ``Meta: Data compression and event detection grand challenge 2024 with sprsound dataset,'' {\em IEEE Data Descriptions}, 2024.

\bibitem{kao2018r}
C.-C. Kao, W.~Wang, M.~Sun, and C.~Wang, ``R-crnn: Region-based convolutional recurrent neural network for audio event detection,'' in {\em Proc. Interspeech 2018}, pp.~1358--1362, 2018.

\bibitem{chaudhary2021automatic}
P.~K. Chaudhary and R.~B. Pachori, ``Automatic diagnosis of glaucoma using two-dimensional fourier-bessel series expansion based empirical wavelet transform,'' {\em Biomedical Signal Processing and Control}, vol.~64, p.~102237, 2021.

\end{thebibliography}


\begin{thebibliography}{10}

\bibitem{world2022world}
W.~H. Organization {\em et~al.}, ``World health statistics 2022: monitoring health for the sdgs, sustainable development goals,'' 2022.

\bibitem{mathers2006projections}
C.~D. Mathers and D.~Loncar, ``Projections of global mortality and burden of disease from 2002 to 2030,'' {\em PLoS medicine}, vol.~3, no.~11, p.~e442, 2006.

\bibitem{harding2020global}
E.~Harding, ``Who global progress report on tuberculosis elimination,'' {\em The Lancet Respiratory Medicine}, vol.~8, no.~1, p.~19, 2020.

\bibitem{forum2017global}
F.~of~International Respiratory~Societies, ``The global impact of respiratory disease--second edition. sheffield, european respiratory society,'' 2017.

\bibitem{bohadana2014fundamentals}
A.~Bohadana, G.~Izbicki, and S.~S. Kraman, ``Fundamentals of lung auscultation,'' {\em New England Journal of Medicine}, vol.~370, no.~8, pp.~744--751, 2014.

\bibitem{bahoura2003new}
M.~Bahoura and C.~Pelletier, ``New parameters for respiratory sound classification,'' in {\em CCECE 2003-Canadian Conference on Electrical and Computer Engineering. Toward a Caring and Humane Technology (Cat. No. 03CH37436)}, vol.~3, pp.~1457--1460, IEEE, 2003.

\bibitem{li2021lungattn}
J.~Li, J.~Yuan, H.~Wang, S.~Liu, Q.~Guo, Y.~Ma, Y.~Li, L.~Zhao, and G.~Wang, ``Lungattn: advanced lung sound classification using attention mechanism with dual tqwt and triple stft spectrogram,'' {\em Physiological Measurement}, vol.~42, no.~10, p.~105006, 2021.

\bibitem{shuvo2020lightweight}
S.~B. Shuvo, S.~N. Ali, S.~I. Swapnil, T.~Hasan, and M.~I.~H. Bhuiyan, ``A lightweight cnn model for detecting respiratory diseases from lung auscultation sounds using emd-cwt-based hybrid scalogram,'' {\em IEEE Journal of Biomedical and Health Informatics}, vol.~25, no.~7, pp.~2595--2603, 2020.

\bibitem{gupta2022classification}
S.~Gupta, M.~Agrawal, and D.~Deepak, ``Classification of auscultation sounds into objective spirometry findings using mvmd and 3d cnn,'' in {\em 2022 National Conference on Communications (NCC)}, pp.~42--47, IEEE, 2022.

\bibitem{xu2021arsc}
L.~Xu, J.~Cheng, J.~Liu, H.~Kuang, F.~Wu, and J.~Wang, ``Arsc-net: Adventitious respiratory sound classification network using parallel paths with channel-spatial attention,'' in {\em 2021 IEEE International Conference on Bioinformatics and Biomedicine (BIBM)}, pp.~1125--1130, IEEE, 2021.

\bibitem{bardou2018lung}
D.~Bardou, K.~Zhang, and S.~M. Ahmad, ``Lung sounds classification using convolutional neural networks,'' {\em Artificial intelligence in medicine}, vol.~88, pp.~58--69, 2018.

\bibitem{chen2019triple}
H.~Chen, X.~Yuan, Z.~Pei, M.~Li, and J.~Li, ``Triple-classification of respiratory sounds using optimized s-transform and deep residual networks,'' {\em IEEE Access}, vol.~7, pp.~32845--32852, 2019.

\bibitem{minami2019automatic}
K.~Minami, H.~Lu, H.~Kim, S.~Mabu, Y.~Hirano, and S.~Kido, ``Automatic classification of large-scale respiratory sound dataset based on convolutional neural network,'' in {\em 2019 19th International Conference on Control, Automation and Systems (ICCAS)}, pp.~804--807, IEEE, 2019.

\bibitem{ma2019lungbrn}
Y.~Ma, X.~Xu, Q.~Yu, Y.~Zhang, Y.~Li, J.~Zhao, and G.~Wang, ``Lungbrn: A smart digital stethoscope for detecting respiratory disease using bi-resnet deep learning algorithm,'' in {\em 2019 IEEE Biomedical Circuits and Systems Conference (BioCAS)}, pp.~1--4, IEEE, 2019.

\bibitem{gupta2021gammatonegram}
S.~Gupta, M.~Agrawal, and D.~Deepak, ``Gammatonegram based triple classification of lung sounds using deep convolutional neural network with transfer learning,'' {\em Biomedical Signal Processing and Control}, vol.~70, p.~102947, 2021.

\bibitem{li2010feature}
S.~Li and Y.~Liu, ``Feature extraction of lung sounds based on bispectrum analysis,'' in {\em 2010 Third International Symposium on Information Processing}, pp.~393--397, IEEE, 2010.

\bibitem{sen2015comparison}
I.~Sen, M.~Saraclar, and Y.~P. Kahya, ``A comparison of svm and gmm-based classifier configurations for diagnostic classification of pulmonary sounds,'' {\em IEEE Transactions on Biomedical Engineering}, vol.~62, no.~7, pp.~1768--1776, 2015.

\bibitem{cinyol2023incorporating}
F.~Cinyol, U.~Baysal, D.~K{\"o}ksal, E.~Babao{\u{g}}lu, and S.~S. Ula{\c{s}}l{\i}, ``Incorporating support vector machine to the classification of respiratory sounds by convolutional neural network,'' {\em Biomedical Signal Processing and Control}, vol.~79, p.~104093, 2023.

\bibitem{shehab2024deep}
S.~A. Shehab, K.~K. Mohammed, A.~Darwish, and A.~E. Hassanien, ``Deep learning and feature fusion-based lung sound recognition model to diagnoses the respiratory diseases,'' {\em Soft Computing}, pp.~1--17, 2024.

\bibitem{topuz2024super}
E.~K. Topuz and Y.~Kaya, ``Super-cough: A super learner-based ensemble machine learning method for detecting disease on cough acoustic signals,'' {\em Biomedical Signal Processing and Control}, vol.~93, p.~106165, 2024.

\bibitem{song2023patch}
W.~Song and J.~Han, ``Patch-level contrastive embedding learning for respiratory sound classification,'' {\em Biomedical Signal Processing and Control}, vol.~80, p.~104338, 2023.

\bibitem{song2021contrastive}
W.~Song, J.~Han, and H.~Song, ``Contrastive embeddind learning method for respiratory sound classification,'' in {\em ICASSP 2021-2021 IEEE International Conference on Acoustics, Speech and Signal Processing (ICASSP)}, pp.~1275--1279, IEEE, 2021.

\bibitem{bae2023patch}
S.~Bae, J.~Kim, W.~Cho, H.~Baek, S.~Son, B.~Lee, C.~Ha, K.~Tae, S.~Kim, and S.~Yun, ``Patch-mix contrastive learning with audio spectrogram transformer on respiratory sound classification,'' 2023.

\bibitem{moummad2023pretraining}
I.~Moummad and N.~Farrugia, ``Pretraining respiratory sound representations using metadata and contrastive learning,'' in {\em 2023 IEEE Workshop on Applications of Signal Processing to Audio and Acoustics (WASPAA)}, pp.~1--5, IEEE, 2023.

\bibitem{hou2023audio}
Y.~Hou, S.~Song, C.~Yu, W.~Wang, and D.~Botteldooren, ``Audio event-relational graph representation learning for acoustic scene classification,'' {\em IEEE signal processing letters}, 2023.

\bibitem{castro2024graph}
A.~E. Castro-Ospina, M.~A. Solarte-Sanchez, L.~S. Vega-Escobar, C.~Isaza, and J.~D. Mart{\'\i}nez-Vargas, ``Graph-based audio classification using pre-trained models and graph neural networks,'' {\em Sensors}, vol.~24, no.~7, p.~2106, 2024.

\bibitem{jiang2024sound}
Y.~Jiang, D.~Guo, L.~Wang, H.~Zhang, H.~Dong, Y.~Qiu, and H.~Zou, ``Sound event detection in traffic scenes based on graph convolutional network to obtain multi-modal information,'' {\em Complex \& Intelligent Systems}, pp.~1--16, 2024.

\bibitem{imoto2020sound}
K.~Imoto and S.~Kyochi, ``Sound event detection utilizing graph laplacian regularization with event co-occurrence,'' {\em IEICE TRANSACTIONS on Information and Systems}, vol.~103, no.~9, pp.~1971--1977, 2020.

\bibitem{wang2022musicyolo}
X.~Wang, B.~Tian, W.~Yang, W.~Xu, and W.~Cheng, ``Musicyolo: A vision-based framework for automatic singing transcription,'' {\em IEEE/ACM Transactions on Audio, Speech, and Language Processing}, vol.~31, pp.~229--241, 2022.

\bibitem{wang2022musicyolo2}
X.~Wang, W.~Xu, W.~Yang, and W.~Cheng, ``Musicyolo: A sight-singing onset/offset detection framework based on object detection instead of spectrum frames,'' in {\em ICASSP 2022-2022 IEEE International Conference on Acoustics, Speech and Signal Processing (ICASSP)}, pp.~396--400, IEEE, 2022.

\bibitem{venkatesh2022you}
S.~Venkatesh, D.~Moffat, and E.~R. Miranda, ``You only hear once: a yolo-like algorithm for audio segmentation and sound event detection,'' {\em Applied Sciences}, vol.~12, no.~7, p.~3293, 2022.

\bibitem{kao2018r}
C.-C. Kao, W.~Wang, M.~Sun, and C.~Wang, ``R-crnn: Region-based convolutional recurrent neural network for audio event detection,'' in {\em Proc. Interspeech 2018}, pp.~1358--1362, 2018.

\bibitem{ye2021sound}
Z.~Ye, X.~Wang, H.~Liu, Y.~Qian, R.~Tao, L.~Yan, and K.~Ouchi, ``Sound event detection transformer: An event-based end-to-end model for sound event detection,'' {\em arXiv preprint arXiv:2110.02011}, 2021.

\bibitem{mesaros2016metrics}
A.~Mesaros, T.~Heittola, and T.~Virtanen, ``Metrics for polyphonic sound event detection,'' {\em Applied Sciences}, vol.~6, no.~6, p.~162, 2016.

\bibitem{ferroni2021improving}
G.~Ferroni, N.~Turpault, J.~Azcarreta, F.~Tuveri, R.~Serizel, {\c{C}}.~Bilen, and S.~Krstulovi{\'c}, ``Improving sound event detection metrics: insights from dcase 2020,'' in {\em ICASSP 2021-2021 IEEE international conference on acoustics, speech and signal processing (ICASSP)}, pp.~631--635, IEEE, 2021.

\bibitem{bilen2020framework}
{\c{C}}.~Bilen, G.~Ferroni, F.~Tuveri, J.~Azcarreta, and S.~Krstulovi{\'c}, ``A framework for the robust evaluation of sound event detection,'' in {\em ICASSP 2020-2020 IEEE International Conference on Acoustics, Speech and Signal Processing (ICASSP)}, pp.~61--65, IEEE, 2020.

\bibitem{lang2020graph}
R.~Lang, R.~Lu, C.~Zhao, H.~Qin, and G.~Liu, ``Graph-based semi-supervised one class support vector machine for detecting abnormal lung sounds,'' {\em Applied Mathematics and Computation}, vol.~364, p.~124487, 2020.

\bibitem{lang2021analysis}
R.~Lang, Y.~Fan, G.~Liu, and G.~Liu, ``Analysis of unlabeled lung sound samples using semi-supervised convolutional neural networks,'' {\em Applied Mathematics and Computation}, vol.~411, p.~126511, 2021.

\bibitem{jacome2019convolutional}
C.~J{\'a}come, J.~Ravn, E.~Holsb{\o}, J.~C. Aviles-Solis, H.~Melbye, and L.~Ailo~Bongo, ``Convolutional neural network for breathing phase detection in lung sounds,'' {\em Sensors}, vol.~19, no.~8, p.~1798, 2019.

\bibitem{hsu2021benchmarking}
F.-S. Hsu, S.-R. Huang, C.-W. Huang, C.-J. Huang, Y.-R. Cheng, C.-C. Chen, J.~Hsiao, C.-W. Chen, L.-C. Chen, Y.-C. Lai, {\em et~al.}, ``Benchmarking of eight recurrent neural network variants for breath phase and adventitious sound detection on a self-developed open-access lung sound database—hf\_lung\_v1,'' {\em PLoS One}, vol.~16, no.~7, p.~e0254134, 2021.

\bibitem{hsu2023dual}
F.-S. Hsu, S.-R. Huang, C.-F. Su, C.-W. Huang, Y.-R. Cheng, C.-C. Chen, C.-Y. Wu, C.-W. Chen, Y.-C. Lai, T.-W. Cheng, {\em et~al.}, ``A dual-purpose deep learning model for auscultated lung and tracheal sound analysis based on mixed set training,'' {\em Biomedical Signal Processing and Control}, vol.~86, p.~105222, 2023.

\bibitem{yeh2024novel}
C.-Y. Yeh, S.-A. Chiu, X.-Y. Deng, and W.-C. Fang, ``A novel ai-inspired method and system implementation for detecting and classifying pediatric respiratory sound events,'' in {\em 2024 IEEE Biomedical Circuits and Systems Conference (BioCAS)}, pp.~1--5, IEEE, 2024.

\bibitem{zhang2024meta}
Q.~Zhang, C.~Chen, S.~Yuan, J.~Zhang, J.~Yuan, H.~Huang, Y.~Zhang, R.~Pan, X.~Jiang, J.~Zhao, {\em et~al.}, ``Meta: Data compression and event detection grand challenge 2024 with sprsound dataset,'' {\em IEEE Data Descriptions}, 2024.

\bibitem{chaudhary2021automatic}
P.~K. Chaudhary and R.~B. Pachori, ``Automatic diagnosis of glaucoma using two-dimensional fourier-bessel series expansion based empirical wavelet transform,'' {\em Biomedical Signal Processing and Control}, vol.~64, p.~102237, 2021.

\end{thebibliography}

\end{document}